\documentclass[aps, pra, a4paper, showpacs, twocolumn, english, 10pt]{revtex4-2}
\usepackage{bbm, amsthm, bm, textcomp, nicefrac, geometry,ragged2e}
\usepackage[dvipsnames]{xcolor}
\usepackage{float}
\usepackage[bbgreekl]{mathbbol}
\usepackage{graphicx, epstopdf, color, verbatim, enumitem,ulem}
\usepackage{pbox,array}
\usepackage{mathrsfs}
\usepackage{multirow}
\usepackage{amssymb}
\usepackage{textgreek}
\usepackage{mathtools}
\usepackage{stackrel}
\usepackage[thinlines]{easytable}
\usepackage{amsmath}
\usepackage[utf8]{inputenc}
\makeatletter
\usepackage{soul}
\usepackage[caption=false]{subfig}
\usepackage{hyperref}
\hypersetup{
    colorlinks = true,
    linkcolor = Purple,
    citecolor = Purple,
    filecolor = Purple,
    urlcolor = Purple,
}
\usepackage[T1]{fontenc}
\usepackage[caption=false]{subfig}
\usepackage{babel}
\usepackage[linesnumbered,ruled]{algorithm2e}

\addto\captionsenglish{}

\newcommand\id{\mathbbm{1}}
\newcommand{\ket}[1]{\left| #1 \right\rangle}
\newcommand{\proj}[1]{\left| #1 \right\rangle\! \left\langle #1 \right|}
\newcommand{\ketbra}[2]{\left| #1\right\rangle\!\left\langle#2\right|}

\newcommand{\bra}[1]{\left\langle #1 \right|}
\newcommand{\bigbra}[1]{\big\langle #1 \big|}

\definecolor{brickred}{rgb}{0.8, 0.0, 0.0}

\begin{document}

\title{Quantum Repeater for \textit{W} States}

\author{Jorge Miguel-Ramiro}
\thanks{These authors contributed equally to this work.}

\author{Ferran Riera-S{\`a}bat}
\thanks{These authors contributed equally to this work.}

\author{Wolfgang D\"ur}

\affiliation{Universit\"at Innsbruck, Institut f\"ur Theoretische Physik, Technikerstra{\ss}e 21a, Innsbruck 6020, Austria}

\date{\today}

\begin{abstract}
W states are a valuable resource for various quantum information tasks, and several protocols to generate them have been proposed and implemented. We introduce a quantum repeater protocol to efficiently distribute three-qubit $W$ states over arbitrary distances in a 2D triangular quantum network with polylogarithmic overhead, thereby enabling these applications between remote parties. The repeater protocol combines two ingredients that we establish: probabilistic entanglement swapping with three copies of three-qubit $W$ states to a single long-distance three-qubit $W$ state, and an improved entanglement purification protocol. The latter not only shows a better performance, but also an enlarged purification regime as compared to previous approaches. We show that the repeater protocol allows one to deal with errors resulting from imperfect channels or state preparation, and noisy operations, and we analyze error thresholds, achievable fidelities and overheads. 
\end{abstract}

\maketitle

\section{Introduction}

Entanglement is the key resource for many quantum information tasks. Different kinds and classes of entanglement have been identified \cite{Dur2000W, Horodecki2009, Walter2016}, and shown to be useful for various applications in quantum information \cite{PhysRevLett.67.661, PhysRevLett.70.1895, PhysRevLett.118.220501, sheng2022367, PhysRevA.59.4249}. The most prominent entangled states include Bell states, Greenberger-Horne-Zeilinger (GHZ) states \cite{Greenberger1990}, graph states \cite{Hein2004, He06}, and the $W$ state \cite{Dur2000W}. For each of these state families, several applications are known. While graph states have found applications in quantum error correction \cite{GottesmanThesis} or as a resource for measurement-based quantum computation \cite{Briegel2001}, GHZ states are, e.g., used in quantum metrology \cite{Giovannetti2011}. The $W$ state has been shown to provide an advantage in secure communication \cite{Joo02, Liu2011, Lip2018, Delocalized2020}, secret voting \cite{Hondt2006}, and multiple other quantum information tasks \cite{Shi2002, Agrawal2006, Choi2010, Ng2014, pirandola2017fundamental}. There exists a multitude of works that discuss how to generate and manipulate $W$ states in different setups, and $W$ states have indeed been generated in several platforms including linear optics \cite{Eibl2004, Sheng2012}, trapped ions \cite{Hffner2005, Cole2021} or superconducting devices \cite{Kang2016}.

Here we consider the distribution of $W$ states over long distances in a quantum network or communication scenario. We introduce a quantum repeater protocol that is capable of efficiently generating $W$ states over arbitrary distances, with an overhead that only grows polylogarithmical with the distance. To this aim, we consider a triangular two-dimensional (2D) quantum network, where noisy, elementary three-qubit $W$ states are probabilistically merged via an entanglement-swapping-type process to form a three-qubit $W$ state over longer distance. The loss in fidelity is compensated by employing an entanglement purification for $W$ states, where out of many noisy copies fewer states with improved fidelity are generated. This procedure is iteratively applied using the resulting states of the previous level as elementary copies. Similar as in the standard quantum repeater for Bell states \cite{Briegel1998, Dur1999, Ladd_2006, RevModPhys.83.33, sheng2013hybrid, azuma2015all, zwerger2016measurement, zwerger2018long, pirandola2017fundamental, pirandola2019end, Azuma2022},
or its 2D variant for GHZ and graph states \cite{Wallnofer2016, Epping2016, Wallnfer2019}, this does not only allow one to overcome errors in state preparation or quantum channels, but also to deal with noise and imperfections in local operations and measurements that are required to perform state merging and entanglement purification. The unique entanglement properties of $W$ states \cite{Dur2000W}, in comparison with, e.g., GHZ or graph states, make this state conceptually further away from bipartite structures, and therefore generalization of bipartite protocols, such as entanglement swapping or purification, presumably requires more challenging and dedicated strategies. 

While many schemes have been discussed on how to enlarge the size of $W$ states to more qubits \cite{Tashima2008,Fujii2011,Bugu2013,Wei2020}, here we introduce a merging or entanglement swapping procedure that generates the same kind of state, i.e., a $W$ state of three qubits--though distributed over a longer distance (see Fig. \ref{fig:Wrepeater}). This is the key to obtaining self-similar structures over longer and longer distances. The protocol succeeds only probabilistically with $p^{\rm succ}=2/3$, but this is no hindrance to getting an efficient scheme to distribute $W$ states over arbitrary distances. 

Additionally, we propose an entanglement purification protocol for $W$ states, a modified strategy of the only existing approach that can distil arbitrarily noisy $W$ states \cite{Miyake2005}, which notably enhances the performance. In particular, our protocol can achieve fidelity arbitrarily close to one, enhanced fidelity improvement, higher success probability, and lower resource overheads. In addition, the purification regime is enlarged, i.e., states with a smaller initial fidelity ($F_0 \approx 0.465$ as compared to $F_0 \approx 0.48$) can be purified. We study the performance of the entanglement swapping and purification procedures, as well as the repeater protocol, and compute achievable fidelities and error thresholds for channels and local operations. We find that local operational noise of up to $2 \%$ in the imperfect implementation of the protocols can be tolerated.

The paper is organized as follows. In Sec. \ref{sec:background} we review concepts and tools we make use of, including protocols to purify $W$ states, and existing quantum repeater protocols for Bell states and graph states. We also describe the error model and the overall setting of the quantum network we consider. In Sec. \ref{sec:swapping} we introduce our entanglement swapping protocol for $W$ states, and discuss the performance of a quantum relay based on it, including also the influence of noise and imperfections. In Sec. \ref{sec:purificationsection} we describe our proposal for an improved entanglement purification of $W$ states, and compare it with the existing approaches. We describe the quantum repeater protocol in Sec. \ref{sec:repeater}, and analyze its properties with respect to achievable fidelities, error thresholds, and overheads. We summarize and conclude in Sec. \ref{sec:conclusion}.     

\section{Background and setting}\label{sec:background}

We review in this section all the concepts and tools we make use of throughout this work.

\subsection{Bell states}
The unit of maximal entanglement between two qubits is given by a Bell state. In particular, we denote the four Bell states that form an orthonormal basis of the two-qubit Hilbert space $\mathcal{H} =\mathbb{C}^{2} \otimes \mathbb{C}^{2}$ \cite{jnielsen2002quantum}, as
\begin{align}
    &\ket{\phi^{\pm}}_{AB}= \frac{1}{\sqrt{2}} \left( \ket{00}_{AB} \pm \ket{11}_{AB} \right), \notag \\
    &\ket{\psi^{\pm}}_{AB}= \frac{1}{\sqrt{2}} \left( \ket{01}_{AB} \pm \ket{10}_{AB} \right).
    \label{eq:bell}
\end{align}
A Bell measurement is defined as a two-qubit measurement with respect to the Bell basis, i.e., the outcomes found are one out of the four Bell states defined above.

When more qubits share entanglement, the features and properties of the states become more complex and diverse. In particular, different entanglement classes arise, containing states that cannot be transformed into states of another class employing local operations \cite{Horodecki2009}. 

\subsection{\textit{W} states}
\label{sec:w}
The three-qubit $W$ state is defined as
\begin{equation}
    \label{eq:W}
    \ket{\text{W}} = \frac{1}{\sqrt{3}} \big( \ket{001} + \ket{010} + \ket{100} \big), 
\end{equation}
and can also be understood as a particular case of a three-particle Dicke state with a single excitation \cite{Dicke54}, whose generalization for a larger number of parties is direct, i.e., $\ket{\text{W}_n} = \sum_{k=1}^n X_k \ket{0}^{\otimes n}/\sqrt{n}$, where $X_k$ is Pauli-$X$ operator acting on qubit $k$.

$W$ states constitute one of the two nonequivalent entanglement classes for three-qubit systems, with the GHZ state being the representative of the other class \cite{Dur2000W}. This implies that a $W$ state cannot be transformed into a GHZ-like state with local operations. In compliance with this property, $W$ states exhibit different entanglement features, with respect to GHZ states, such as entanglement robustness under losses, that make them unique and particularly promising for different quantum applications.

The challenges of generalizing bipartite protocols (such as entanglement purification or swapping) for $W$ states, in opposition to, e.g., GHZ states, can be better understood by attending to the entanglement properties of both classes of states. Consider a three-qubit GHZ of the form $\ket{\text{GHZ}} = \big( \ket{000} + \ket{111} \big)/\sqrt{2}$, up to local unitaries. The entanglement of each bipartition, i.e., the entanglement of each of the qubits with respect to the rest, which can be easily computed by studying the entropy of entanglement of the reduced density operators \cite{jnielsen2002quantum}, is exactly one ebit (the amount of entanglement contained in a Bell state). A three-party GHZ state can be deterministically transformed into a Bell state between any two constituents by simply measuring the third in the Pauli-$X$ basis. On the other hand, the entanglement structure of the $W$ state, Eq. \eqref{eq:W}, tells us that the entanglement of any qubit with respect to the rest is always strictly less than one. Although this provides certain advantages, such that the aforementioned robustness under noise and losses, it makes it impossible to achieve a deterministic transformation into one of the Bell states between two predetermined parties, a process that can only succeed with some probability \cite{Fortescue2007, Fortescue2008}. This makes GHZ-like states closer in a certain sense to bipartite entanglement, a feature which is evidenced by the more direct generalization of many protocols and tools from Bell to GHZ states. Such straightforward extensions are not usually found with $W$ states, which require dedicated approaches. 

Substantial efforts have been focused on successfully preparing \cite{Eibl2004, Grfe2014} and increasing the size of $W$ states, by fusion \cite{zdemir2011, Bugu2013, Wei2020}, i.e., merging two $W$ states into a bigger one, and expansion techniques \cite{Tashima2008}, i.e., adding qubits to enlarge a $W$ state, for which different experimental realizations have been proven \cite{Thasima2010}. 

\subsection{\textit{W}-state entanglement purification protocol} \label{sec:wpurification}

Entanglement purification protocols (EPPs) \cite{Dur2007} refer to the set of techniques and approaches that allow one to increase the fidelity of entangled states using local operations. From an ensemble of noisy copies, an EPP consists of local joint manipulation of the states in such a way that fewer copies, but with enhanced fidelity, are obtained. The fidelity of some state $\rho$ is the typical figure of merit used to evaluate the protocols, and it is given by
\begin{equation*}
    F = \left\langle \psi \right| \rho  \left|\psi \right\rangle,
\end{equation*}
with respect to the pure target state $\ket{\psi}$.

EPPs are utilized in repeater schemes as a fundamental building block. When aiming to construct long communication links, the detrimental effects of noise become worse with the distance. EPPs provide a tool to recursively recover the fidelity of entangled copies, by sacrificing others, at designed intermediate stations, such that, when combined with entanglement swapping (see below), the aforementioned high-quality long-distance connections can be achieved.

Different approaches have been proposed for bipartite entanglement purification \cite{Bennett96, Deutsch96, PhysRevA.105.062418, yan2023advances}, and generalizations to multipartite entanglement, including different entanglement structures, such as, e.g., GHZ-like states \cite{Dur2003Pur}, or graph states \cite{Dur2003Pur, Aschauer2005}, have been proposed. 

However, EPP approaches for $W$-like states have drawn less interest. To the best of our knowledge, there exists only a single proposal that achieves purification of arbitrary noisy $W$ states \cite{Miyake2005}, which is based on stabilizer measurements on three noisy copies that iteratively correct for most kinds of errors. In the following, we denote this purification protocol as ``stabilizer EPP''. Alternative approaches have been proposed that correct only specific kinds of noise affecting a $W$ state, such as, e.g., amplitude damping \cite{Sun2012, Huang2014}.

We briefly review here the stabilizer EPP strategy and propose in Sec. \ref{sec:purificationsection} modifications of the approach that significantly increase the protocol working performance, and therefore the efficiency of the $W$-state repeater.

The stabilizer EPP introduced in Ref. \cite{Miyake2005} consists in alternating between two recurrence subroutines, denoted as $\mathcal{P}$ and $\bar{\mathcal{P}}$. Each routine takes as input three identical copies of a three-qubit noisy $W$-state, operates jointly (but still locally) on the three copies and probabilistically provides a single state with increased fidelity, i.e., $\mathcal{P}/\bar{\mathcal{P}}\!: \,\rho_{ABC}^{\otimes 3} \to \rho'_{ABC}$. In each iteration of the protocol, subroutine $\mathcal{P}$ or $\bar{\mathcal{P}}$ is heuristically implemented depending on which one provides higher fidelity.

\textit{Subroutine $\mathcal{P}$.} Each party ($A$, $B$ and $C$) performs a projective measurement of the form $\left\{ M^{000}, M^{001}, M^{010}, M^{100} \right\}$ on their three qubits (corresponding to the three different copies), where
\begin{equation}
    \label{eq:MM}
    M^{ijk} = \proj{W^{ijk}} + X^{\otimes 3} \proj{W^{ijk}} X^{\otimes 3},
\end{equation}
and $\ket{W^{ijk}}_{123} = (Z_2 X_3 + Z_1 X_2 + X_1 Z_3) \ket{ijk}_{123}/\sqrt{3}$. The remaining three-qubit subsystems are mapped into a single-qubit one following  $\left\{ \ket{W^{001}}, \ket{W^{010}}, \ket{W^{100}} \right\} \rightarrow \ket{0}$ and $\left\{ \ket{W^{011}}, \ket{W^{110}}, \ket{W^{101}} \right\} \rightarrow \ket{1}$.

In case the three parties obtain the same outcome and, as long as the outcome is not $M^{000}$, the protocol iteration succeeds and the remaining $W$ state exhibits an increased fidelity. Otherwise, the state is discarded.

\textit{Subroutine $\bar{\mathcal{P}}$.} In this dual routine, each party first applies a local basis change $V$, which leaves unchanged $\ket{000}$ and $\ket{111}$, but exchanges $\ket{001} \leftrightarrow \ket{110}$, $\ket{010} \leftrightarrow \ket{101}$ and $\ket{100} \leftrightarrow \ket{011}$, i.e., $\rho_{ABC}^{\otimes 3} \to V_A V_B V_C \, \rho_{ABC}^{\otimes 3} \, V_A V_B V_C$. Then, each party performs a projective measurement given by $\left\{ \bar{M}^{000}, \id - \bar{M}^{000}\right\}$ on their three qubits, where $\bar{M}^{000} = \Lambda^\dagger M^{000}\Lambda$ [see Eq. \eqref{eq:MM}], with $\Lambda = H_1 H_2 H_3 \, \text{SWAP}_{1,3}$. In case the three parties obtain the outcome $\bar{M}^{000}$, the protocol iteration succeeds and the fidelity of the final state is eventually increased. Otherwise, the state is discarded. 

By selectively choosing subroutines $\mathcal{ P }$ or $\bar{\mathcal{ P }}$ an (almost) perfect $W$ state can be distilled. In Ref. \cite{Miyake2005} a numerical analysis of the protocol is provided, showing an upper bound on the achievable fidelity of $F=0.999$. We show in Sec. \ref{sec:purificationsection} how to overcome this limitation, by substituting subroutine $\bar{\mathcal{ P }}$, also leading to increased performance and success probability.

\subsection{Quantum repeater}
\label{sec:basicrepeater}

Quantum repeaters \cite{Briegel1998, Dur1999, Ladd_2006, RevModPhys.83.33, sheng2013hybrid, azuma2015all, zwerger2016measurement, zwerger2018long, pirandola2019end} are fundamental communication tools that allow for long-range quantum communication overcoming the detrimental effect of noise and decoherence that naturally affect qubits and entanglement when physically distributed. Such a goal is achieved by combining entanglement swapping \cite{Briegel1998} with entanglement purification \cite{Dur1999} in dedicated intermediate stations. Quantum repeaters have been proven and analyzed experimentally \cite{Pan1998, Li2019, Kruty2022} and constitute a basic building block for viable quantum networks \cite{Kimble2008, rev1, Azuma_2021}. 

The simplest repeater construction is the bipartite one, where Bell pairs are established. In order to overcome the exponential reduction in success probability and fidelity when transmitting quantum information over longer distances directly, one splits up the channel into smaller segments and places intermediate repeater stations. Bipartite entangled states are distributed between neighboring repeater stations in such a way that each repeater station shares an ensemble of noisy copies with the previous and the next one. Subsequently, a Bell measurement in a repeater station between qubits of differently directed Bell copies, a process also denoted as entanglement swapping, directly leads to a Bell state of larger distance, however with reduced fidelity. One can then use EPP strategies (see Sec. \ref{sec:wpurification}) to obtain fewer copies with improved fidelity. EPPs operate locally on multiple copies and in general succeed probabilistically. One then faces exactly the same situation as in the beginning but with elementary Bell states of enlarged distance and same (or higher) fidelity. One can then apply the same procedure in a nested fashion, doubling the distance of resulting Bell pairs in each step while maintaining their fidelity. This allows one to achieve high-fidelity Bell states over arbitrary distances, with an overhead in resources that scales only polylogarithmic with distance \cite{Briegel1998, Dur1999}. Notice that it is important to run the protocol in a nested way, as a sequential application of entanglement swapping and entanglement purification still leads to an exponential scaling of resources. One may also start with an EPP step if the fidelity of the initially generated elementary copies is not large enough.    

This basic repeater functionality has been generalized for multipartite settings \cite{Wallnofer2016, Epping2016, Wallnfer2019}, where the desired long-distant connections involve more than two parties. In particular, extensions with GHZ \cite{Wallnofer2016, Kuzmin2019}, cluster states \cite{Wallnofer2016} and graph states \cite{Epping2016} have been proposed, such that an architecture for establishing long-distance entangled states of that kind is built in a 2D fashion, and analyzed for different realizations.

As mentioned above, we focus here on developing a repeater scheme for a different kind of multipartite entanglement, i.e., $W$ states.
\begin{figure}
    \centering
    \includegraphics[width=0.99\columnwidth]{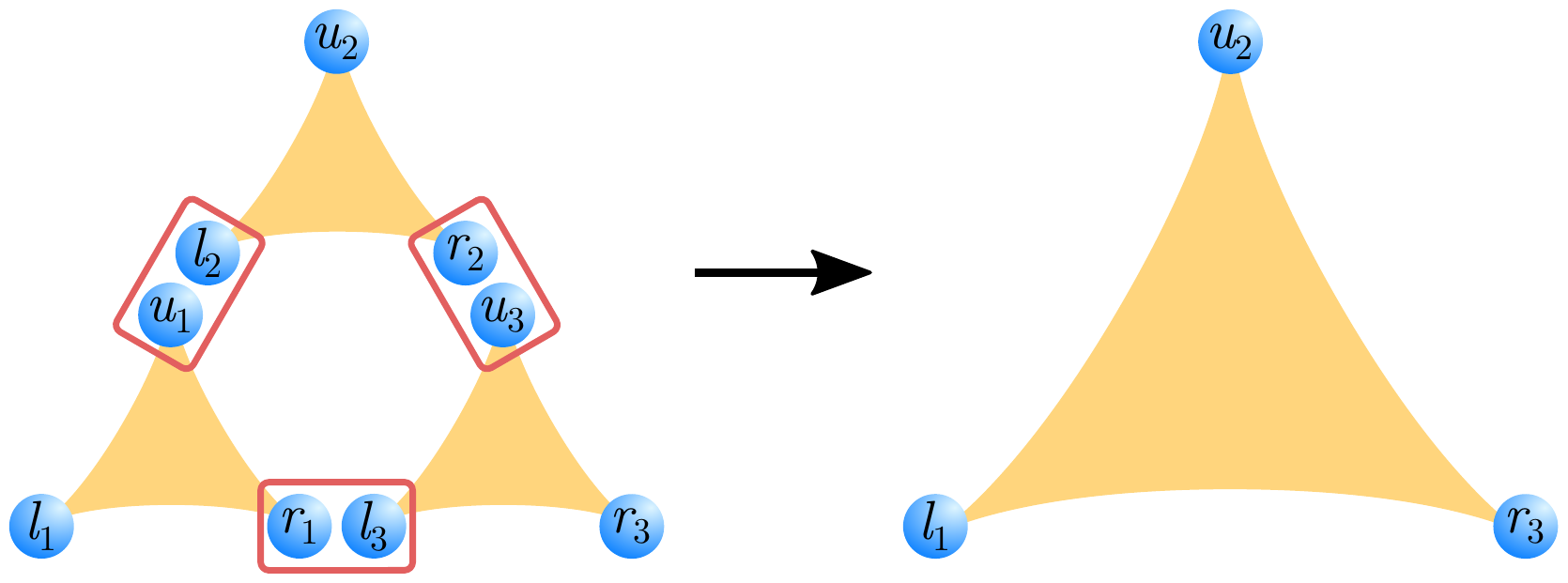}
    \caption{\label{fig:Wrepeater} Basic illustration of the general idea of a quantum repeater for three-qubits $W$ states.}
\end{figure}

\subsection{Setting and noise model}
\label{sec:noise model}

We consider a typical quantum communication scenario, where some source generates perfect three-qubit $W$ states, and distributes each qubit to spatially separated stations. During the distribution, the qubits are subjected to noise and decoherence. We model this process by applying local depolarizing noise to each of the qubits with parameter ``\textit{q}'', of the form

\begin{equation}
\label{eq:depolarizing}
    \mathcal{D} \left(\rho\right) = q \, \rho + \frac{1-q}{4} \sum_{k=0}^3 \sigma_k \, \rho \, \sigma_k,
\end{equation}
where $\sigma_i = \{\mathbb{1}, X, Z, Y \}$ are the Pauli matrices. Depolarizing noise constitutes the worst local noisy case \cite{jnielsen2002quantum}, since all the information is completely lost with probability $(1-q)$, and therefore is a common choice to analyze how noise affects the performance of communication protocols \cite{jnielsen2002quantum}.

Several copies of the noisy $W$ states are then stored in the stations, which locally manipulate their qubits of different $W$ states to implement the repeater protocol we propose in this work, which is comprised of entanglement purification (if needed) and entanglement swapping. We consider the case where the operations involved in the protocols are not perfect, and we model this by introducing additional local depolarizing noise to each qubit, Eq. \eqref{eq:depolarizing}, with parameter ``\textit{p}'', before performing the ideal operation required for an entanglement swapping or EPP step. We denote this source of noise and imperfection as ``operational noise''.

In this way, a recursive process can be constructed, that allows for a recursive distribution of high-fidelity $W$ states over long distances. In the following, we introduce and explain in detail the entanglement swapping and purification protocols required for a viable repeater process, and analyze the performance of the approach under different situations. 
 
\section{Entanglement swapping for \textit{W} states}
\label{sec:swapping}

We introduce a method to merge three three-qubit $W$ states into a larger three-qubit $W$ state, by locally manipulating the states in a pairwise manner at three different repeater stations, in the same spirit as the so-called 2D quantum repeaters \cite{Wallnofer2016}. The protocol succeeds with probability $2/3$ and provides a tool to build repeater stations for a viable distribution of such kinds of states. 

\begin{figure}
    \centering
    \includegraphics[width=0.75\columnwidth]{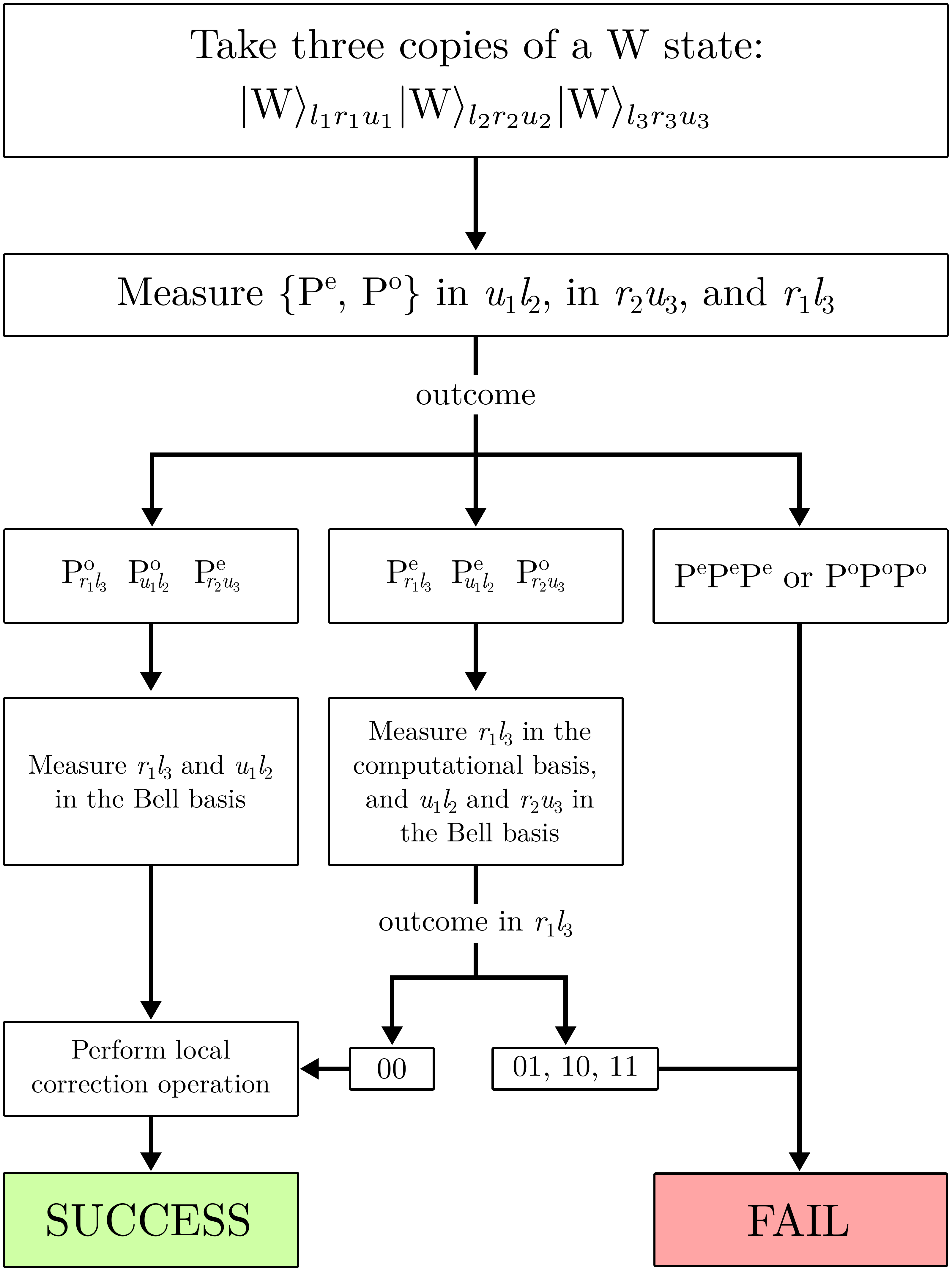}
    \caption{\label{fig:tree} Schematic treelike representation of the entanglement swapping protocol for a $W$ state. }
\end{figure}

We consider three identical copies of a $W$ state, where we label the qubits as $\ket{W}_{l_1 r_1 u_1} \ket{W}_{l_2 r_2 u_2}\ket{W}_{l_3 r_3 u_3}$, where $l_i, r_i, u_i$ indicate the left, right and upper qubit of the $i$th states respectively (see Fig. \ref{fig:Wrepeater}). In practice, we find three copies of a mixed state $\rho^W$ where, without loss of generality, we assume that the component $\proj{W}$, Eq.~\eqref{eq:W}, is dominant. The protocol comprises the following steps, see also Fig. \ref{fig:tree}.
\begin{itemize}
    \item[0.] The states are distributed and arranged as depicted in Fig. \ref{fig:Wrepeater}, such that one can operate jointly on each pair of qubits $\{l_1,u_2\}$, $\{r_1,u_3\}$ and $\{r_2,l_3\}$ in each so-called repeater station. Note the symmetry degeneracy in the arranging of the qubits.
    
    \item[1.] At each repeater station, the pair of qubits is measured with respect to a two-outcome projective measurement given by $\{ P^o, P^e \}$, where
\begin{equation}
\begin{aligned}
\label{eq:povm}
    P^e & = \proj{00} + \proj{11}, \\
    P^o & = \proj{01} + \proj{10}.
\end{aligned}
\end{equation}
    The results of the measurements are classically communicated to the other stations. Observe how the measurement projects both qubits into a two-dimensional subspace, such that the qubits remain entangled with the rest in a certain way. 
 
    \item[2.] Depending on the combination of outcomes found in the previous step, the repeater stations proceed as follows. Note the results apply for all permutations of outcomes.
    
    \item[2.1.] Outcome $P^o \, P^o \, P^e$. A measurement in the Bell basis is performed in the states where $P^o$ was obtained in step 1, while no operation is required in the station where $P^e$ was found. The protocol succeeds when the outcomes of the Bell measurements are $\ket{\psi^+}$ or $\ket{\psi^-}$, Eq.~\eqref{eq:bell} (not necessarily coinciding). Local Pauli corrections, depicted in Table \ref{Table:corrections} for each case, are finally applied.
    
    \item[2.2.] Outcome $P^e \, P^e \, P^o$. One of the two pairs where $P^e$ is obtained is measured in the computational basis, while the other two pairs are measured in the Bell basis. The protocol succeeds if, for the qubits measured in the computational basis the outcome $\ket{00}$ is found, and the Bell measurements of the $P^e$ and $P^o$ pairs of qubits give $\ket{\phi^\pm}$ and $\ket{\psi^\pm}$, respectively. Local Pauli unitary corrections are again applied (see Table \ref{Table:corrections}).
    
    \item[2.3.] Outcome $P^o \, P^o \, P^o$ or $P^e \, P^e \, P^e$. The protocol fails.
\end{itemize}
In a noiseless operation of this repeater protocol, each $i$th favorable outcome succeeds with probability $p_i = 1/9$, leading to an overall protocol success probability
\begin{equation*}
     p^{\text{succ}} = \frac{2}{3}.
\end{equation*}
Importantly, as analyzed later, the repeater protocol succeeds with approximately similar probability, independently of the operational noise or the form of the initial $\rho^W$ states.

\renewcommand{\arraystretch}{1.5}
\begin{table}
\begin{tabular}{|c|c|c|} \hline
\multicolumn{2}{|c|}{Measurement Outcomes} & $\;\;$ Correction operation $\;\;$ \\ \hline \hline
\multirow{4}{*}{$\text{P}_{r_1 l_3}^{\rm o} \text{P}_{u_1 l_2}^{\rm o} \text{P}_{r_2 u_3}^{\rm e}$} & $\psi_{r_1 l_3}^+ \, \psi_{u_1 l_2}^+$ & none \\
                                                             & $\psi_{r_1 l_3}^+ \, \psi_{u_1 l_2}^-$ & $Z_{u_2}$           \\
                                                             & $\psi_{r_1 l_3}^- \, \psi_{u_1 l_2}^+$ & $Z_{r_3}$           \\
                                                             & $\psi_{r_1 l_3}^- \, \psi_{u_1 l_2}^-$ & $Z_{l_1}$           \\ \hline
\multirow{4}{*}{$\text{P}_{r_1 l_3}^{\rm e} \text{P}_{u_1 l_2}^{\rm e} \text{P}_{r_2 u_3}^{\rm o}$} & $P_{r_1 l_3}^{00} \, \phi_{u_1 l_2}^+ \, \psi_{r_2 u_3}^+$ & $X_{l_1}$ \\
                                                             & $P_{r_1 l_3}^{00} \, \phi_{u_1 l_2}^+ \, \psi_{r_2 u_3}^+$ & $Y_{l_1} Z_{u_2}$ \\
                                                             & $P_{r_1 l_3}^{00} \, \phi_{u_1 l_2}^- \, \psi_{r_2 u_3}^+$ & $Y_{l_1}$         \\
                                                             & $\, P_{r_1 l_3}^{00} \, \phi_{u_1 l_2}^- \, \psi_{r_2 u_3}^+$ & $X_{l_1} Z_{u_2} \,$ \\ \hline                   
\end{tabular}
\caption{\label{Table:corrections} Correction operations depending on the different outcomes in the entanglement swapping protocol for $W$ states.}
\end{table}

\subsection{Quantum relay for $W$ states: resource analysis}

We evaluate here the performance of the entanglement swapping protocol in a repeater context, without entanglement purification, memories, and channel decoherence. This approach is also known as a quantum relay \cite{Collins2005, Varnava2016}, which allows one to overcome exponentially dropping success probability of generation and merging processes. Generating elementary copies in optical setups typically have a certain success probability, either because the source produces states probabilistically (e.g., using some parametric down-conversion process), or because of photon-loss errors in channels. Already in a 1D setup, if each elementary copy is produced with probability $p$, the success probability to obtain all $N$ states among the chain and successfully merge them is given by $p^N$. This can be overcome if multiple elementary copies are generated, and the ones where the process is successful are merged with ones from the neighboring link. On average, one needs $1/p$ elementary copies to have one available for merging, and (quasi) deterministically produce one long-distance entangled pair.

In our case, not only the generation process of elementary states may be probabilistic, but also the merging process itself. The effect, however, is similar, and one can overcome exponential scaling by operating only on states where the previous merging process was successful. This requires some (limited) quantum memory, or at least the possibility to selectively operate on different copies. Here we consider a two-dimensional version of a quantum relay, where three elementary $W$ states are merged to form a long-distance one, thereby growing in distance in two dimensions. Importantly, we need to consider that entanglement swapping is performed in a nested way, i.e., that the distance of copies is doubled in each step.

Observe that in the way the $W$-state repeater is conceived, the process can be iterated in a parallel way in the different stations, making the total resources a relevant figure of merit to be evaluated.

We start our analysis by computing the number of resources $R$, i.e., the expected number of $W$ states, required to distribute a single $W$ state over a distance $L$ in the noiseless case. We assume that elementary $W$ states are distributed over a distance $\ell$, and then merged using entanglement swapping, thereby doubling the distance in each step. In the next round, only states where the entanglement swapping process was successful are considered and selected. A single iteration takes as input three $W$ states over a distance $x$ and with probability $p^{\text{succ}}$ generates a $W$ state over a distance $2x$, where the distance is enlarged in two dimensions simultaneously as we deal with a 2D setting. Therefore, the number of resources to establish a $W$ state over a distance $L$ is given by $R_L = 3 \, R_{L/2} / p^{\text{succ}}$. As we assume $R_\ell = 1$, to establish a $W$ state over a distance $L = 2^n \ell$ the number of resources, i.e., the total number of elementary copies, is given by
\begin{equation*}
    R_L = \left( \frac{9}{2} \right)^n = \left( \frac{L}{\ell} \right)^{2 \log_2 (3)-1} \approx \left( \frac{L}{\ell} \right)^{2.17}.
\end{equation*}
The number of copies per segment is given by $R_{\rm seg} = \left(L/\ell\right)^{\log_2 (3)-1} \approx \left(L/\ell\right)^{0.59}$.

\subsection{Entanglement swapping for \textit{W} states: Perfect operations}

Next, we continue our analyses by concentrating on the $W$ entanglement-swapping process with channel noise. We investigate the impact of noise from the generation of states, and from the quantum channels through which qubits are sent to the stations. The merging of three noisy $W$ states leads to a larger $W$ state which is however subjected to a higher amount of noise, with in general an altered form. 

We identify operational thresholds based on the minimum fidelity required to purify a $W$ state. This fidelity is slightly higher than the one found in Ref. \cite{Miyake2005} for random $W$ mixed states. In our case, numerical analysis indicates that any state $\rho^W$ coming from a repeater process can be purified as long as its fidelity is larger than $F_0 \sim 0.465$ (see Sec. \ref{sec:purificationsection}).

In addition, we make use of different parameters to evaluate the protocol's performance. The distance is given by $2^n$, where $n$ is the number of iterative rounds. This can be understood from the fact that, after each swapping round, the distance between any two parties sharing a $W$ state is doubled. Observe also that after $n$ iterations, $3^n$ states are used to generate the larger target three-party $W$ state.

\begin{figure}
    \centering
    \subfloat[]{\includegraphics[width=0.48\columnwidth]{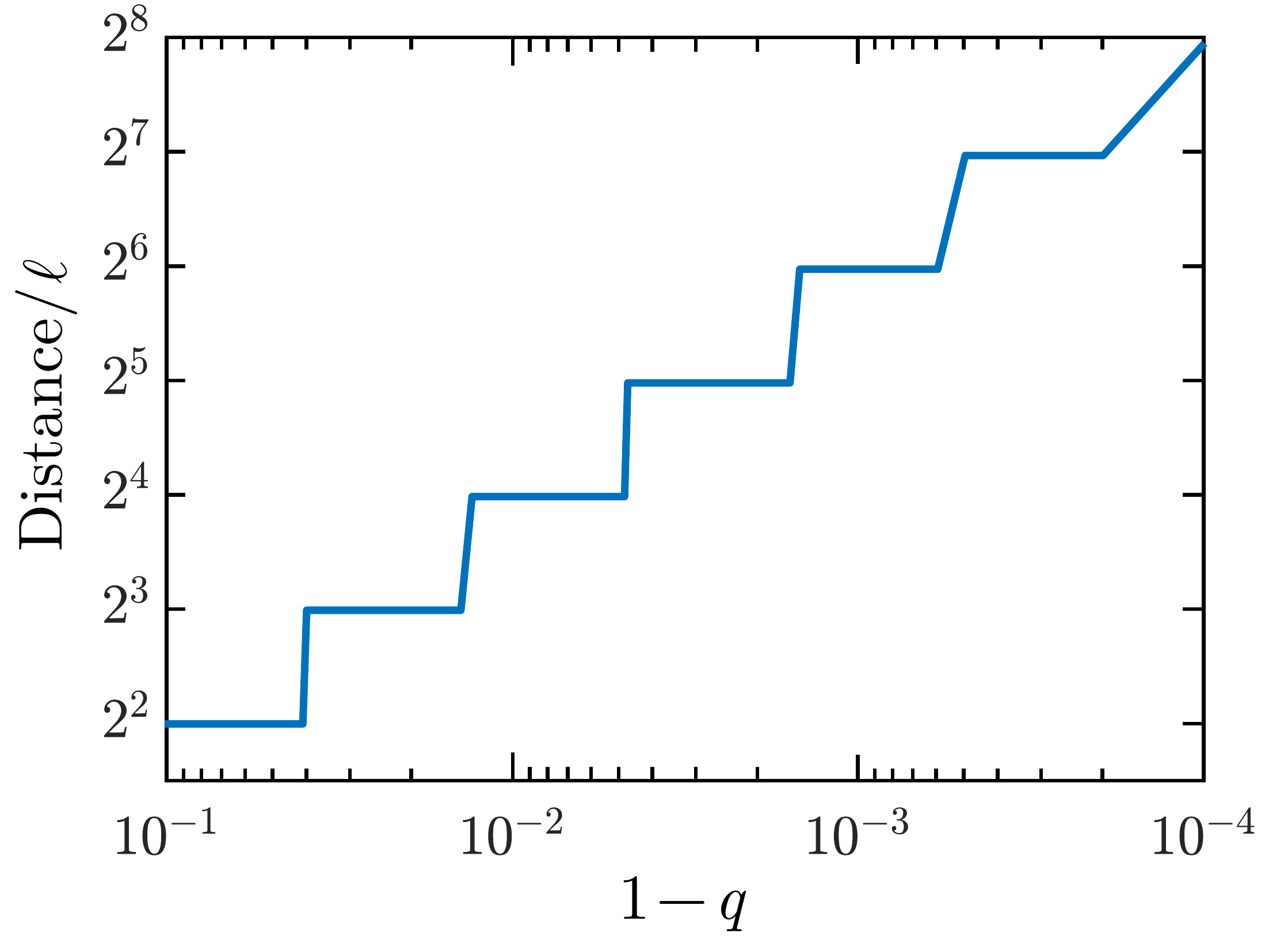}\label{fig:repeaternoisy:a}}
    \subfloat[]{\includegraphics[width=0.48\columnwidth]{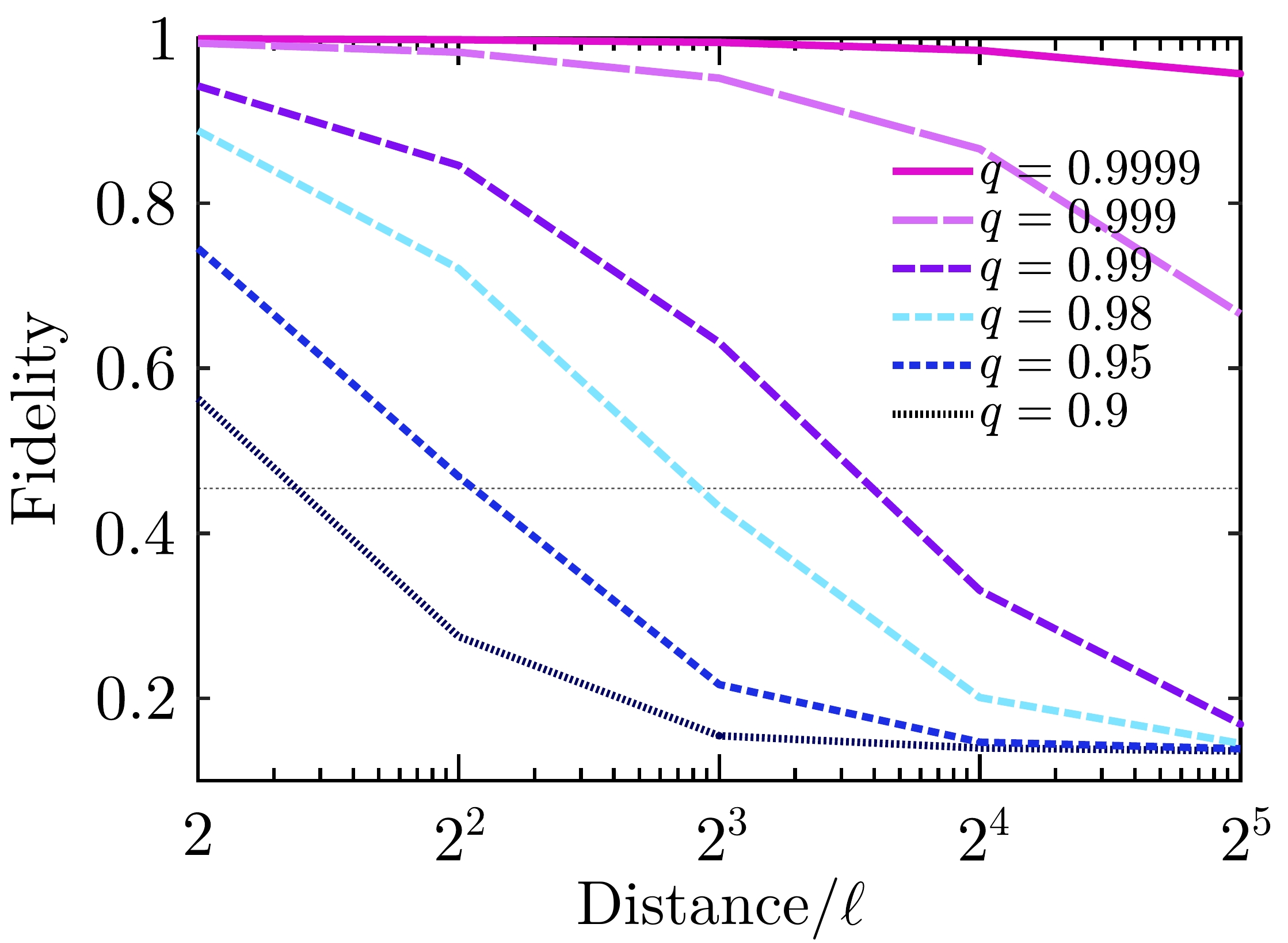}\label{fig:repeaternoisy:b}} \\
    \subfloat[]{\includegraphics[width=0.48\columnwidth]{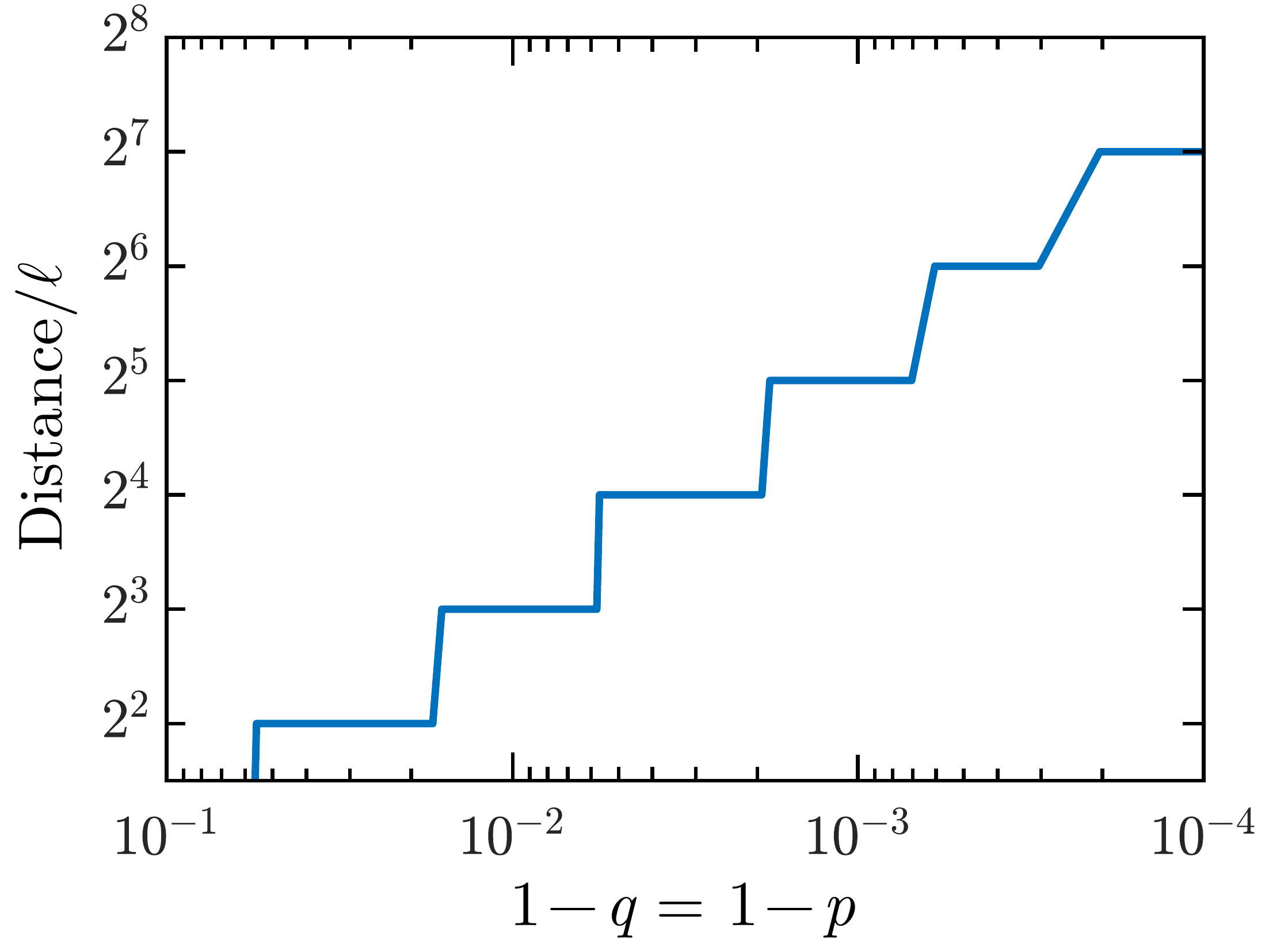}\label{fig:repeaternoisy:c}}
    \subfloat[]{\includegraphics[width=0.48\columnwidth]{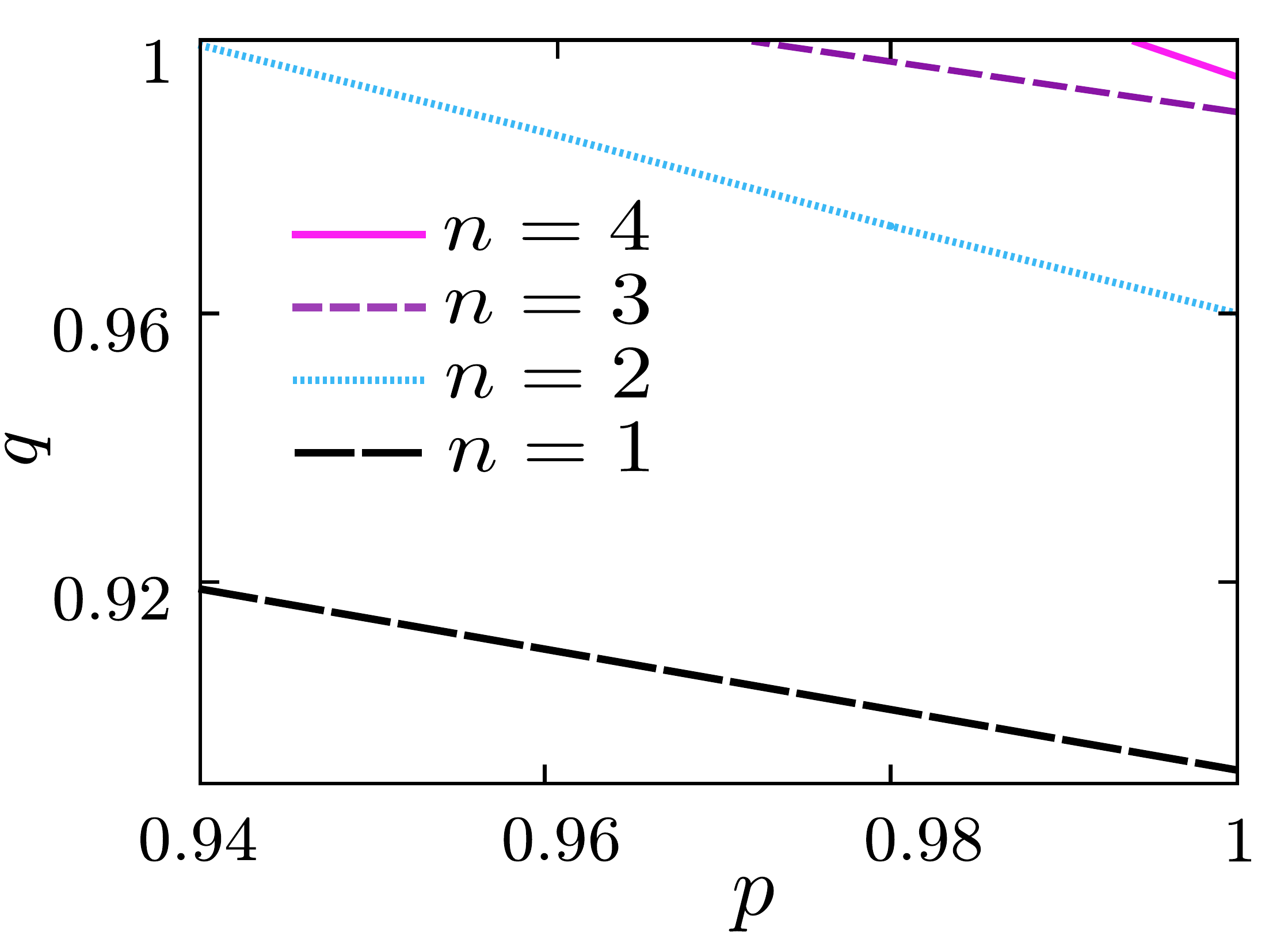}\label{fig:repeaternoisy:d}}
    \caption{\label{fig:repeaternoisy} Performance of the entanglement swapping approach for $W$ states with perfect (a-b) and imperfect (c-d) local operations and measurements specified by error parameter $p$, and channel noise specified by parameter $q$. (a) Achievable distance $2^n \ell$ (where $n$ is the number of consecutive rounds, and the elementary $W$ states are distributed over a distance $\ell$) before the fidelity drops below the purification threshold $F_0 \sim 0.465$. (b) Fidelity of the obtained states as a function of the distance. (c) Achievable distance $2^n \ell$ with imperfect operations ($p=q$) before the fidelity drops below the purification threshold. (d) Maximum tolerable channel noise $q$ as a function of the operation noise $p$ for different distances. }
\end{figure}

Figure \ref{fig:repeaternoisy} shows the entanglement swapping performance as a function of the channel noise parameter and the distance with perfect and noisy operations. The noise model used is the one introduced in Sec. \ref{sec:noise model}, where initial copies are affected by a depolarizing noise of strength $q$, and operations by a depolarizing channel of strength $p$. In Figs. \ref{fig:repeaternoisy:a} and \ref{fig:repeaternoisy:c} we observe the reachable distance before the fidelity drops below the purification threshold in perfect and imperfect operation mode, respectively. In Fig. \ref{fig:repeaternoisy:b} we show how the fidelity decreases with the distance. Finally in Fig. \ref{fig:repeaternoisy:d} we plot the maximum tolerable channel noise $q$ as a function of the operation noise $p$ for different distances.

All figure results presented are obtained by proper simulations of the whole protocol.

\subsection{Entanglement swapping for \textit{W} states: Imperfect operations}

Here we investigate the effect of imperfect operations during the entanglement swapping routine. We model this effect by introducing local depolarizing noise in every qubit before the ideal implementation of the operations. This simplified model allows one to illustrate the impact of noise in the same way as done for 1D \cite{Briegel1998, Dur1999} and 2D \cite{Wallnofer2016, Epping2016, Wallnfer2019} standard quantum repeaters, however with the crucial difference that no entanglement purification is considered.

The performance of the entanglement swapping under these circumstances is analyzed in Figs. \ref{fig:repeaternoisy:c} and \ref{fig:repeaternoisy:d}. Figure \ref{fig:repeaternoisy:c} shows the reachable distance when the operational noise is comparable to the channel noise, where we observe a moderate efficiency reduction compared to the operational noiseless case. On the other hand, Fig. \ref{fig:repeaternoisy:d} analyzes the threshold \textit{q} value that corresponds to the maximum acceptable channel noise for a given operational noise and for different numbers of rounds. 

\section{Improved entanglement purification for $W$ states} \label{sec:purificationsection}

In Sec. \ref{sec:wpurification} we have reviewed the only existing EPP \cite{Miyake2005} that has been proposed to purify arbitrarily noisy $W$ states. We propose here a modification of that protocol that offers several immediate advantages: lower required fidelity, significantly better performance (higher fidelities are reached with fewer steps), enhanced success probability, and unbounded achievable fidelity.  

\begin{figure}
    \centering
    \subfloat[]{\includegraphics[width=0.48\columnwidth]{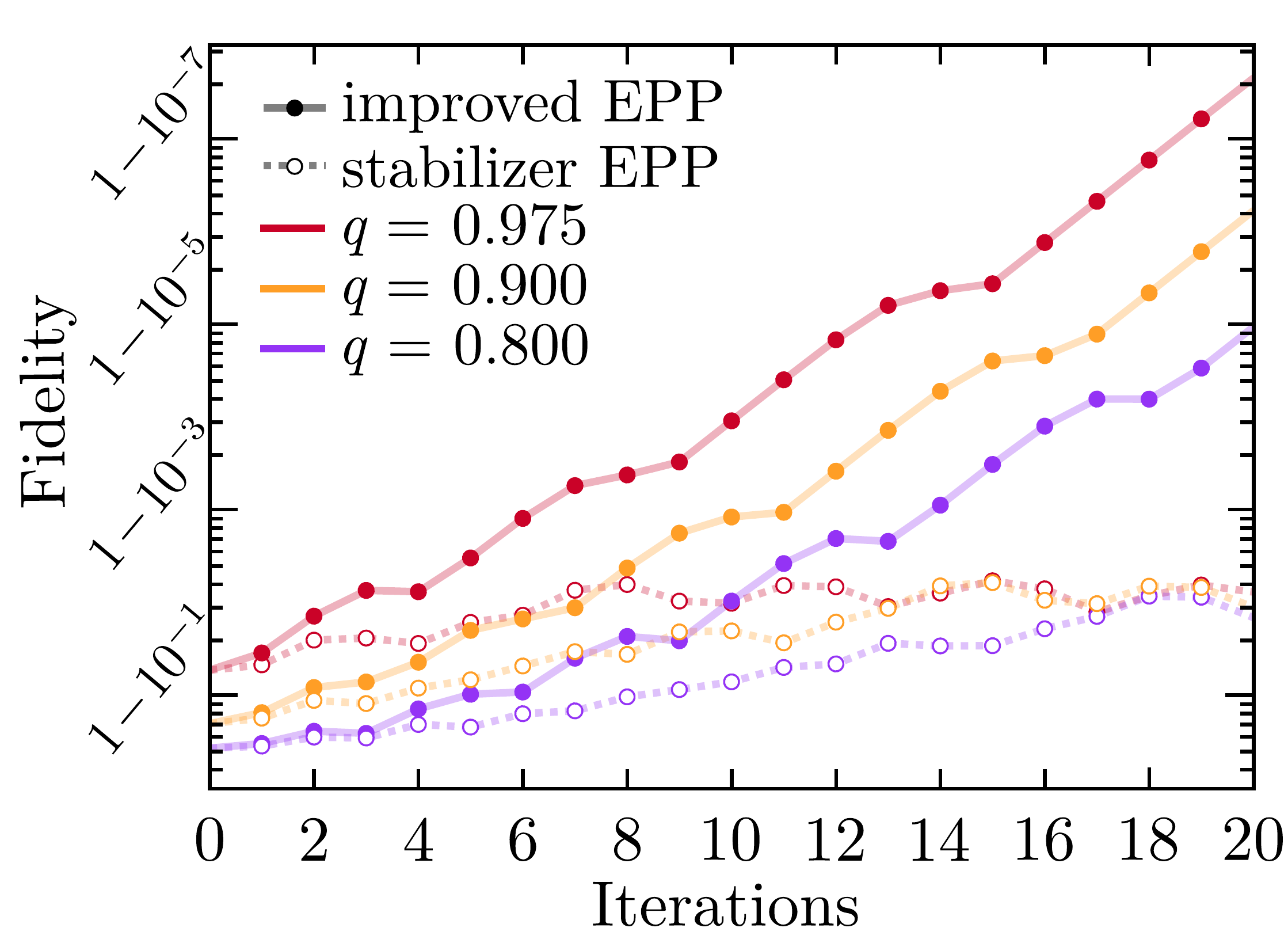}\label{fig:purification:a}}
    \subfloat[]{\includegraphics[width=0.48\columnwidth]{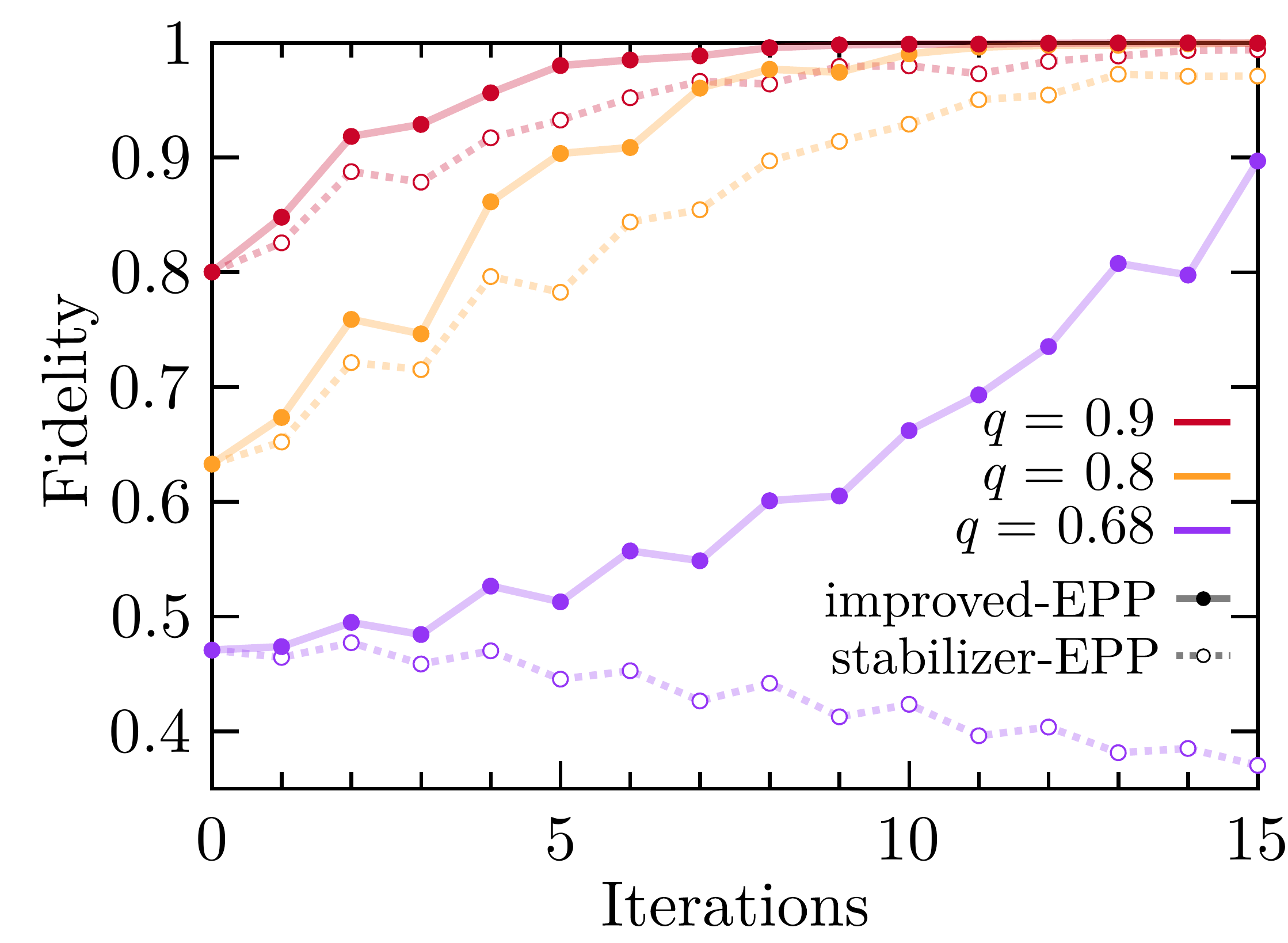}\label{fig:purification:b}} \\
    \subfloat[]{\includegraphics[width=0.48\columnwidth]{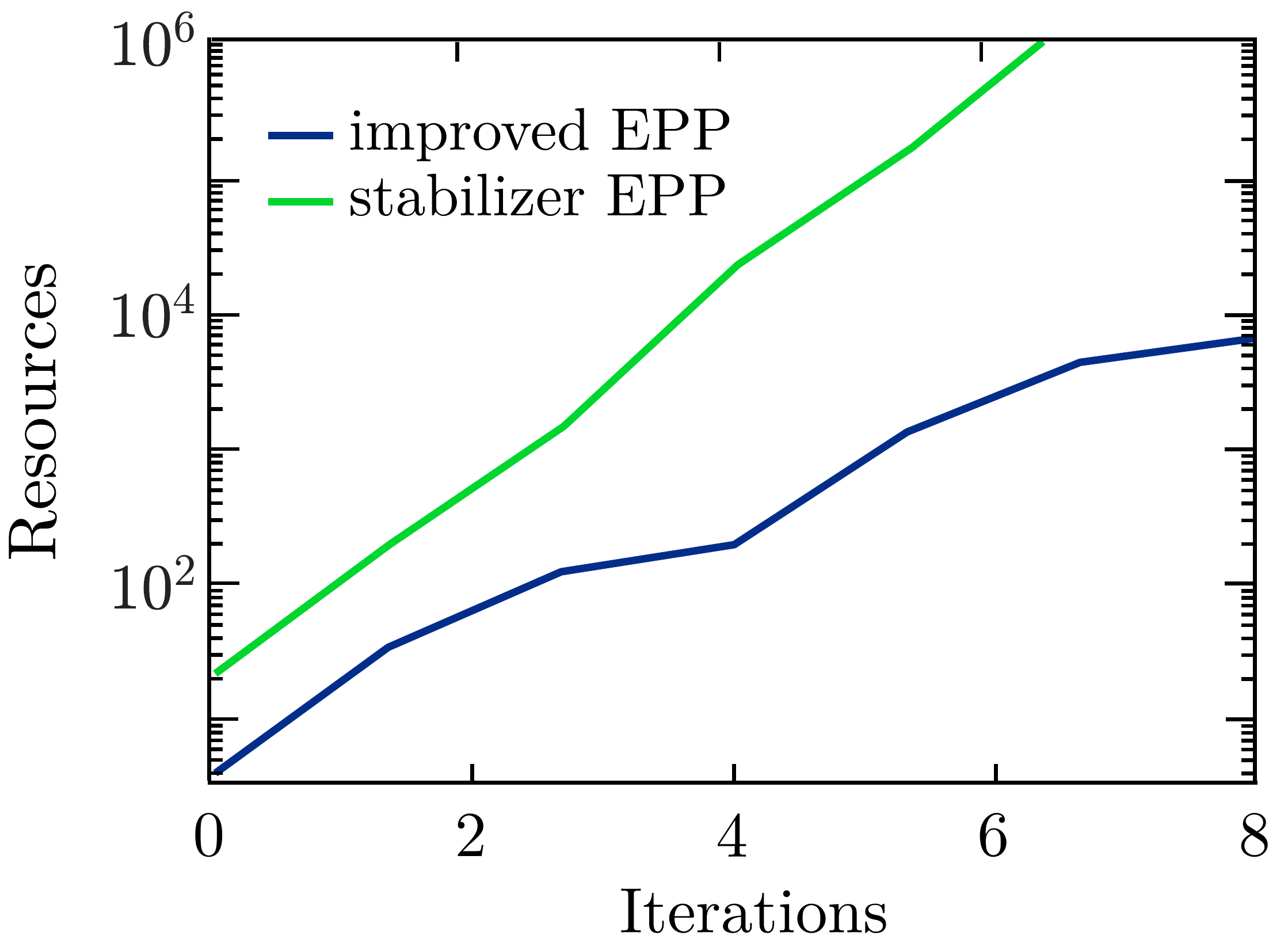}\label{fig:purification:c}}
    \subfloat[]{\includegraphics[width=0.48\columnwidth]{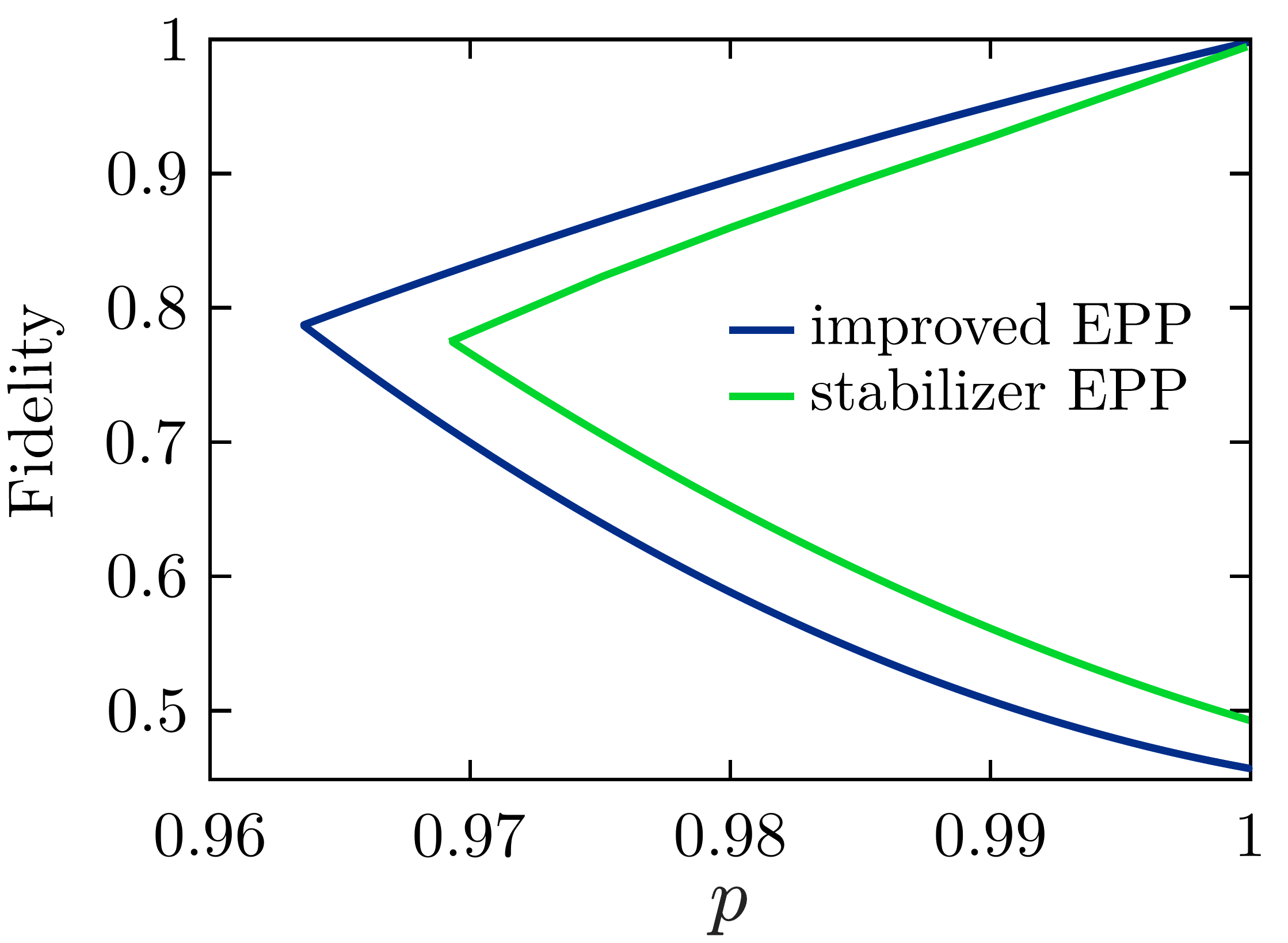}\label{fig:purification:d}}
    \caption{\label{fig:purification}Comparison of EPP performance between our improved EPP proposal and the stabilizer EPP strategy from Ref. \cite{Aschauer2005}. (a-b) Fidelity evolution in a logarithmic (a) and linear (b) scaling as a function of the number of iterations of the EPP for $W$ states with different initial fidelities. (c) The accumulative amount of resources required for each protocol for initial $W$ states with local depolarizing noise of strength $q=0.9$. (d) Maximum achievable and minimum required fidelities as a function of the operational noise described by local depolarizing noise with strength $p$ acting before each iteration.    }
\end{figure}

The stabilizer EPP approach of Ref. \cite{Miyake2005}, see also Sec. \ref{sec:wpurification}, exhibits certain limitations mainly due to the dual subroutine $\bar{\mathcal{ P }}$. We identify and propose an alternative subprotocol, substituting the routine $\bar{\mathcal{ P }}$ by $\mathcal{ P }'$, that overcomes these limitations. This improved EPP version hence consists of two recurrence subroutines,  $\mathcal{ P }$ and  $\mathcal{ P }'$, which are again iteratively applied, acting on three or two copies respectively, probabilistically providing a state of increased fidelity, i.e., $\mathcal{ P }'\!:  \, \rho_{\rm ABC}^{\otimes 2} \to \rho_{\rm ABC}'$.

\textit{Subroutine $\mathcal{ P }$.} Identical to subprotocol $\mathcal{ P }$ from Ref. \cite{Miyake2005}. See  Sec. \ref{sec:wpurification}.

\textit{Subroutine $\mathcal{ P }'$.} Only two copies of the noisy $W$ state are manipulated. Each party ($A$, $B$, $C$) performs the two-outcome measurement projecting into even and odd subspace, described by projectors ${P}^e, P^o$ [see Eq. \eqref{eq:povm}]. Only in case all three parties find outcome $P^e$, the protocol iteration succeeds. After performing the mapping (relabeling) $\ket{00} \rightarrow \ket{0}$ and $\ket{11} \rightarrow \ket{1}$ into a single-qubit subsystem at each party, we obtain again a three-qubit system, but with increased fidelity. If any other outcome is found, the state is discarded. 

In every purification step, the subroutine  $\mathcal{P}$ or $\mathcal{P}'$ that provides the higher fidelity is applied. 
\begin{figure*}[ht]
    \centering
    \subfloat[]{\includegraphics[width=0.69\columnwidth]{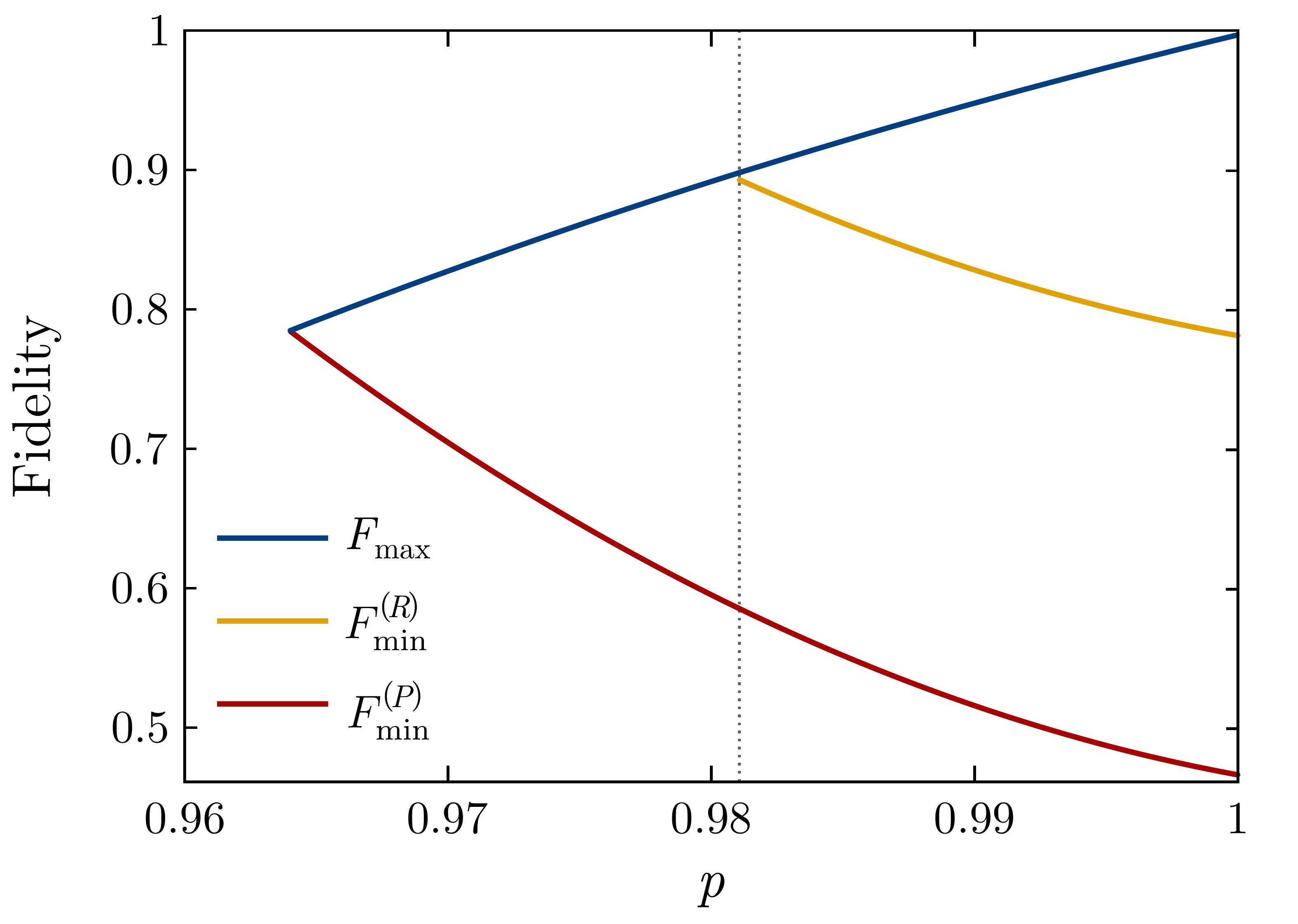}\label{fig:relay:a}}
    \subfloat[]{\includegraphics[width=0.69\columnwidth]{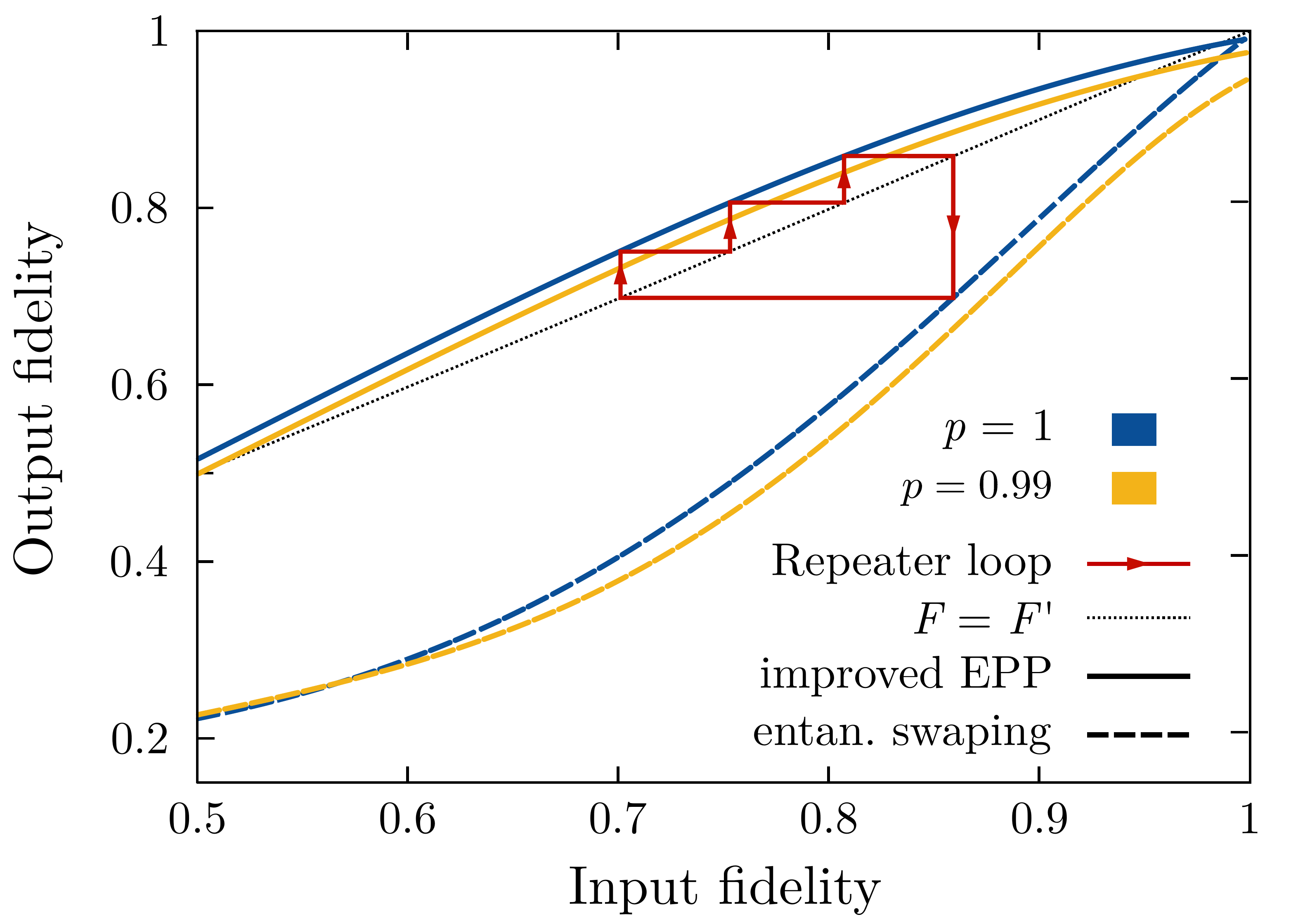}\label{fig:relay:b}}
    \subfloat[]{\includegraphics[width=0.69\columnwidth]{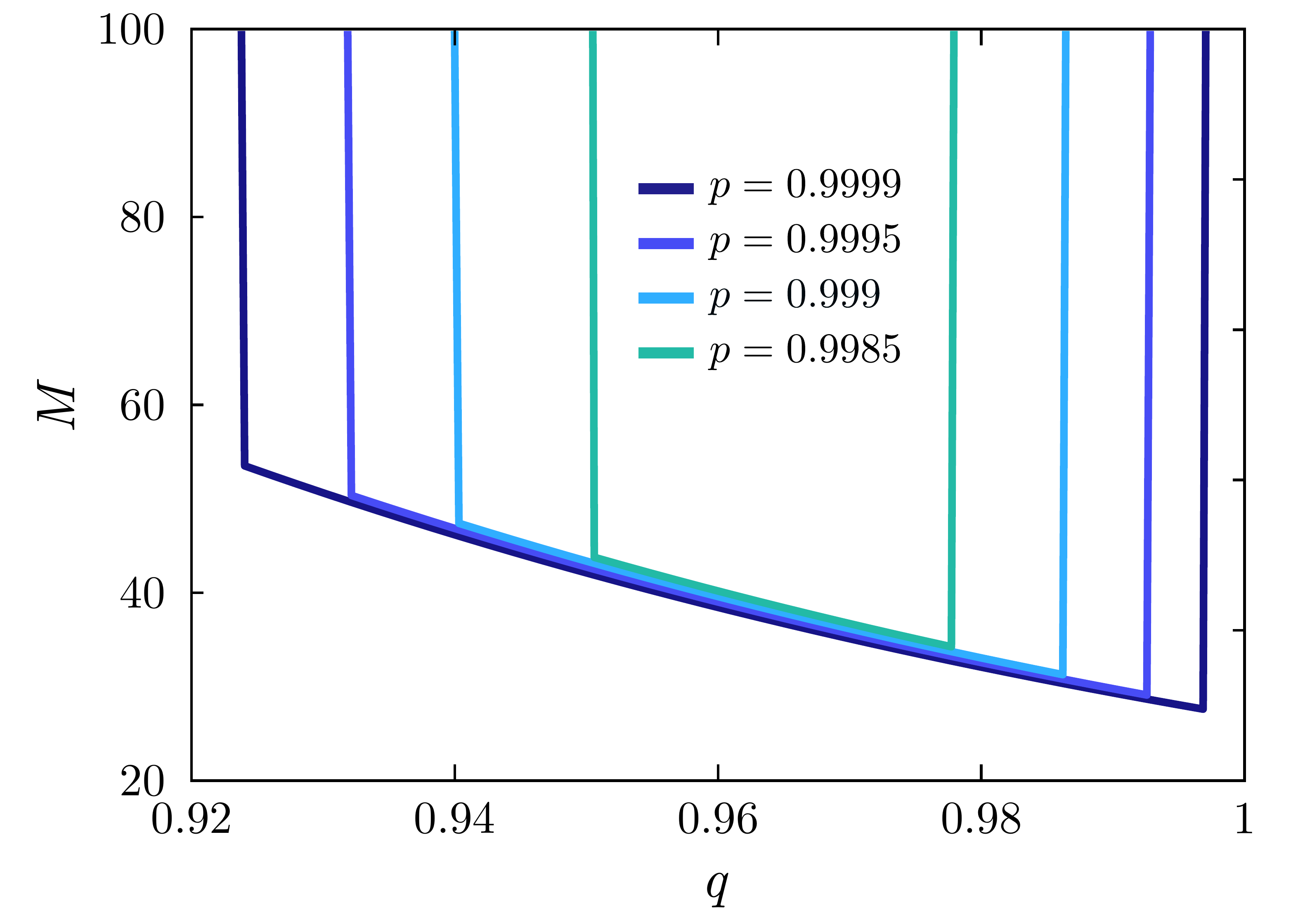}\label{fig:relay:c}}
    \caption{\label{fig:relay} Performance of the improved EPP and the quantum repeater for $W$ states. (a) Minimum required and maximum achievable fidelity, as a function of the operational noise, i.e., local depolarizing noise with parameter $p$ acting on all qubits before each swapping of purification iteration. The different curves correspond to maximal reachable fidelity using EPP ($F_{\rm max}$, blue) and minimal required fidelity for EPP ($F^{(P)}$, red), and the minimal required fidelity to perform a repeater round, i.e., entanglement swapping and EPP ($F^{(R)}$, yellow). (b) Repeater loop that shows the enhancement of fidelity after one round of EPP (solid), and fidelity reduction due to entanglement swapping  (dashed), with ideal (blue) and $1 \%$ (yellow) operational noise. (c) Amount of local resources, $M$, required to maintain the repeater operation for different local noises.}
\end{figure*}
In Fig. \ref{fig:purification}, we provide numerical evidence of the performance enhancement of this modified protocol with respect to the stabilizer EPP \cite{Miyake2005}. In Figs. \ref{fig:purification:a} and \ref{fig:purification:b}, we show the fidelity after each iteration for noisy-W states subjected to local depolarizing noise as input. Observe that (1) for any number of iterations a higher fidelity is obtained and (2) unlike the stabilizer EPP, our improved protocol can provide purified states with fidelity arbitrarily close to one. Moreover, in Fig. \ref{fig:purification:c} we show the number of required resources for a certain number of purification steps is notably improved because of a larger success probability, and because subroutine $\mathcal{ P }'$ operates only on two states. This implies that fewer states need to be discarded, and one obtains a higher yield and efficiency. The operating regime under local operations is also enhanced with our improved EPP, as shown in Fig. \ref{fig:purification:d}, where the maximum achievable and the minimum required fidelities are plotted as a function of the local depolarizing strength ($q$) that affects the states before each iteration.  
Finally, we remark that our improved EPP strategy can purify $W$ states affected by local depolarizing noise with fidelity above $F_0 \approx 0.465$ (see in Fig. \ref{fig:purification:b} the initial fidelity from which the purification protocol no longer succeeds), overcoming the threshold of the stabilizer EPP approach, i.e., $F_0 \approx 0.48$ \cite{Miyake2005}.

We can then make use of the modified EPP to improve the performance of the $W$ state quantum repeater, which we analyze in detail in the following section.

\section{Performance of the quantum repeater for \textit{W} states}
\label{sec:repeater}

We finally consider in this section the whole repeater protocol for $W$ states, which includes the entanglement purification and swapping strategies introduced before. We analyze the performance and find operational thresholds of the protocol under imperfect operations that are modeled by applying local depolarizing noise with parameter ``$p$'', Eq. \eqref{eq:depolarizing}, in each qubit before each swapping round and each purification iteration. Noise on elementary $W$ states is described by depolarizing noise acting on each qubit, with parameter ``$q$'', modeling imperfect state preparation and channel noise. 

Figure \ref{fig:relay:a} shows the operational thresholds of the protocols, i.e., minimum required and maximum achievable fidelity, as a function of the local operation noise. One can directly see how, on one hand, the EPP introduced in Sec. \ref{sec:purificationsection} can tolerate around $3.5 \%$ of noise, whereas this threshold gets reduced to slightly less than $2 \%$ when entanglement swapping (Sec. \ref{sec:swapping}) is involved. The latter curve, $F_{\rm min}^{(R)}$ assumes initial states with fidelity given by the maximum reachable fidelity of the EPP, $F_{\rm max}$, which are then combined via entanglement swapping and should still have a fidelity larger than the minimum required fidelity for purification, $F^{(P)}$. Notice that the threshold for required fidelity of initial elementary states is given by $F^{(P)}$, however, the threshold for the repeater protocol is given by the crossing point of $F_{\rm max}$ and $F_{\rm min}^{(R)}$, $p_{\rm min} \approx 0.981$.

Figure \ref{fig:relay:b} shows the repeater loop for connecting $W$ states. The fidelity increment due to the improved EPP (solid lines), and the fidelity reduction due to entanglement swapping (dashed lines) are shown for perfect local operations, and noisy local operations with $1 \%$ of local operational noise. The initial fidelity before each iteration is given by the dotted diagonal. After the swapping process, the fidelity drops to the dashed lines and, in case it is above the purification threshold, it is again increased via purification in several steps, recovering the initial fidelity and completing one repeater round. Similar qualitative behaviour is observed when compared to, e.g., bipartite repeater strategies \cite{Briegel1998, Dur1999}.

Finally, Fig. \ref{fig:relay:c} refers to the number of local resources ($M$) required as a function of the operational fidelity for different local noise scenarios. When comparing these results with bipartite \cite{Briegel1998, Dur1999} or the different multipartite \cite{Wallnofer2016, Epping2016, Wallnfer2019} repeater approaches, one can see the performance of the $W$ state repeater is lower, i.e., more resources are required. One could expect such a lower performance given the nature of the $W$ state (see discussion in Sec. \ref{sec:w}). We nevertheless stress that, with the entanglement swapping and improved purification protocols introduced in this work, one obtains an efficient quantum repeater protocol for $W$ states that can also deal with noise in local operations.

Direct comparison with other multipartite repeater approaches (e.g., Ref. \cite{Wallnofer2016}) is possible. Several of the results we present in this work correspond to the analogue analyses shown in those works. One can observe that the $W$-state repeater exhibits lower performance rates, including worse fidelity thresholds, higher resource requirements, and lower efficiency under operational noise. The differences mainly arise from the purification protocol, since the entanglement swapping approaches differ only in the $2/3$ success probability in the $W$-state case. GHZ states (and graph states in general) have been extensively studied and multiple purification approaches have been proposed. In contrast, $W$-state purification protocols have received less attention and, despite the improvements we present in this paper, we hope that our work can trigger further research in that direction. As discussed in Sec. \ref{sec:w},  fundamental differences of entanglement properties with respect to e.g. GHZ or 2D cluster states, make $W$ states particularly complicated to deal with, however leading to unique features.

\section{Summary and outlook} \label{sec:conclusion}

In this work, we have introduced the first proposal for a quantum repeater of $W$ states, an important and unique class of multipartite entangled states with various applications. We have proposed entanglement swapping and entanglement purification protocols, which are the two main ingredients for a viable quantum repeater, analyzing their performance in ideal and noisy operational regimes. We have demonstrated that our modified EPP has improved performance, including higher efficiency and larger purification range, thereby leading to an efficient repeater protocol with polylogarithmic scaling of resources with the distance.

However, the performance of quantum repeaters is intrinsically linked to the performance of entanglement swapping and entanglement purification protocols. While we believe that the success probability of entanglement swapping for $W$ states cannot be increased to unity, there might be further room for improvement regarding entanglement purification for $W$ states. Despite the enhancement our protocols show over previous approaches, it still seems that these protocols are less efficient than protocols for Bell states and graph states. In particular, existing EPPs for $W$ states are all recurrence protocols, while it is known that hashing and breeding protocols can have a significantly higher yield \cite{Dur2007, zwerger2018long, Wallnfer2019, ferranPRL, ferranPRA}. It would be interesting to develop such improved protocols for $W$ states. 

Nevertheless, we hope that our results can open the doors for further improvements and generalizations. Preliminary results show how similar approaches as the entanglement swapping we introduce here, can be applied for $W$ states of an arbitrary number of qubits, however, it is not clear how to optimize this strategy and generalize it for other Dicke states with different numbers of excitations. Similarly, generalizations of entanglement purification for arbitrarily large $W$ states, and Dicke states in general, and that offer good performance, are required.

\section*{Acknowledgments}
This work was supported by the Austrian Science Fund (FWF) through
Projects No. P36009-N and No. P36010-N. Finanziert von der Europ\"aischen Union - NextGenerationEU.

\renewcommand\appendixname{Appendix}
\appendix

\section{Details of the \textit{W}-state entanglement swapping protocol}

In this section, we provide additional equations for analyzing and reproducing the results of the entanglement swapping protocol shown in Fig. \ref{fig:repeaternoisy} and explained in Sec. \ref{sec:swapping}.

First, we recall the ``even'' and ``odd'' projectors are given by $P^e = \proj{00} + \proj{11}$ and $P^o = \proj{01} + \proj{10}$, and the four Bell states by $\ket{\phi^{\pm}} = \left( \ket{00} \pm \ket{11} \right)/\sqrt{2}$ and $\ket{\psi^{\pm}} = \left( \ket{01} \pm \ket{10} \right)/\sqrt{2}$.

The entanglement swapping protocol starts with three copies of a $W$ state distributed over a distance $L$, and probabilistically outputs a single $W$ state over a distance $2L$, i.e., 
\begin{equation*}
\begin{aligned}
    \left(\rho^L\right)^{\otimes 3} & = \proj{W}_{l_1 u_1 r_1} \otimes \proj{W}_{l_2 u_2 r_2} \otimes \proj{W}_{l_3 u_3 r_3} \\ 
    & \mapsto \rho^{2L} = \proj{W}_{l_1 u_2 r_3}.
\end{aligned}
\end{equation*}
The protocol consists of the steps shown in Sec.~\ref{sec:swapping}. In case in the first step we obtain output $P^o_{r_1l_3} P^o_{u_1l_2} P^e_{r_2 u_3}$, we can proceed with the second step and the output state is given by $O^{(j)}\rho^L O^{(j)}$, where
\begin{equation*}
\begin{aligned}
    & O^{(1)} = \proj{\psi^+}_{r_1 l_3} \proj{\psi^+}_{u_1 l_2} P^e_{r_2 u_3}, \\
    & O^{(2)} = \proj{\psi^+}_{r_1 l_3} \proj{\psi^-}_{u_1 l_2} P^e_{r_2 u_3}, \\
    & O^{(3)} = \proj{\psi^-}_{r_1 l_3} \proj{\psi^+}_{u_1 l_2} P^e_{r_2 u_3}, \\
    & O^{(4)} = \proj{\psi^-}_{r_1 l_3} \proj{\psi^-}_{u_1 l_2} P^e_{r_2 u_3}.
\end{aligned}
\end{equation*}
On the other hand, if in the first step, we obtain $P^e_{r_1l_3} P^e_{u_1l_2} P^o_{r_2 u_3}$, we then measure $r_1l_3$ in the computational basis and $u_1l_2$ and $r_2u_3$ in the Bell basis. The projectors corresponding to the success outcomes are given by
\begin{equation*}
\begin{aligned}
    & O^{(5)} = \proj{00}_{r_1 l_3} \proj{\phi^+}_{u_1 l_2} \proj{\psi^+}_{r_2 u_3}, \\
    & O^{(6)} = \proj{00}_{r_1 l_3} \proj{\phi^+}_{u_1 l_2} \proj{\psi^-}_{r_2 u_3}, \\
    & O^{(7)} = \proj{00}_{r_1 l_3} \proj{\phi^-}_{u_1 l_2} \proj{\psi^+}_{r_2 u_3}, \\
    & O^{(8)} = \proj{00}_{r_1 l_3} \proj{\phi^-}_{u_1 l_2} \proj{\psi^-}_{r_2 u_3}.
\end{aligned}
\end{equation*}
If we obtain the output of projector $O^{(j)}$, the correction operation we need to implement to the resulting state is given by $C^{(j)}$, where
\begin{equation*}
\begin{aligned}
    & C^{(1)} = \id_{l_1 u_2 r_3}, \qquad & & C^{(2)} = \id_{l_1 r_3} Z_{u_2}, \\
    & C^{(3)} = \id_{l_1 u_2} Z_{r_3}, \qquad & & C^{(4)} = Z_{l_1} \id_{u_2 r_3}, \\
    & C^{(5)} = X_{l_1} \id_{u_2 r_3}, \qquad & & C^{(6)} = Y_{l_1} Z_{u_2} \id_{r_3}, \\
    & C^{(7)} = Y_{l_1} \id_{u_2 r_3}, \qquad & & C^{(8)} = X_{l_1} Z_{u_2} \id_{r_3}.
\end{aligned}
\end{equation*}
Therefore, the total success probability of the protocol reads 
\begin{equation*}
    p^{\text{succ}} = 3 \sum_{j=1}^8 \text{tr} \! \left[ \text{O}^{(j)} \left( \rho^L\right)^{\otimes 3} \right] = \frac{2}{3},
\end{equation*}
where the factor 3 on the left side takes into account all permutations for the outcome in the first step, i.e., $P^{o/e}_{r_1l_3} P^{o/e}_{u_1l_2} P^{e/o}_{r_2 u_3}$, $P^{o/e}_{r_1l_3} P^{e/o}_{u_1l_2} P^{o/e}_{r_2 u_3}$ and $P^{e/o}_{r_1l_3} P^{o/e}_{u_1l_2} P^{o/e}_{r_2 u_3}$. 

In the noisy case, the input state consists of three copies of a noisy $W$ state, i.e., $F = \langle W|\rho^L| W\rangle < 1$. In this case, the output state is given by
\begin{equation}
    \label{app:ep:swap}
    \rho^{2L} = \sum_{j=1}^{8} \frac{\text{tr}_{\neg l_1u_2r_3} \! \Big[ C^{(j)} O^{(j)} \left( \rho^L\right)^{\otimes 3} O^{(j) \dagger} C^{(j)}\Big]}{\text{tr} \! \left[ O^{(j)} \left( \rho^L\right)^{\otimes 3} \right] },
\end{equation}
and the success probability is given by
\begin{equation}
    \label{app:ep:swapps}
    p^{\text{succ}} = 3 \sum_{j=1}^8 \text{tr} \! \left[ O^{(j)} \left(\rho^L\right)^{\otimes 3} \right].
\end{equation}
Starting from elementary $W$ states over a distance $\ell$ and iterating the entangling swapping protocol $n$ times, one can prepare a $W$ state over a distance $L = 2^n \ell$. However, if the elementary $W$ states are noisy, the fidelity of the states decreases after each interaction, i.e., $\langle W | \rho^{2L} | W \rangle < \langle W |\rho^L| W \rangle$.

In Fig.~\ref{fig:repeaternoisy} we consider copies of a $W$ state over a distance $\ell$ affected by local depolarizing noise, i.e., $\rho^\ell = D^{\otimes 3}_q (\proj{W})$, where $D_q^{\otimes 3} = \{ \sqrt{q_i q_j q_k} \, \sigma_i \otimes \sigma_j \otimes \sigma_k \}_{i,j,k=0}^3$ and $q_0 = q$ and $q_{1,2,3} = (1-q) / 4$. In Figs.~\ref{fig:repeaternoisy:c} and \ref{fig:repeaternoisy:d} we also consider noisy operations, meaning that before each iteration the input states are affected by local depolarizing noise, i.e., in Eqs. \ref{app:ep:swap} and \ref{app:ep:swapps} we substitute $\rho^L$ by $D_p^{\otimes 3}\left(\rho^L\right)$.

\section{Details of the \textit{W}-state entanglement purification protocol}

In this section, we provide detailed equations for analyzing the stabilizer EPP and the improved EPP introduced in Sec. \ref{sec:purificationsection} (see also Fig. \ref{fig:purification}). Again, we first recall a few definitions. 

The $W$ basis is given by given by
\begin{equation*}
    \ket{W^{ijk}}_{123} = \frac{1}{\sqrt{3}} (Z_2 X_3 + Z_1 X_2 + X_1 Z_3) \ket{ijk}_{123},
\end{equation*}
for $i,j,k \in \{0,1\}$, and the stabilizer projectors by 
\begin{equation*}
    M^{ijk} = \proj{W^{ijk}} + X^{\otimes 3} \proj{W^{ijk}} X^{\otimes 3}.
\end{equation*}

In order to analyze the purification protocols, given a three-qubit input state, $\rho_{k-1} \in \mathcal{H}_\text{A} \otimes \mathcal{H}_\text{B} \otimes \mathcal{H}_\text{C}$, we need to compute the output state for each subroutine $\varrho_k$, its fidelity $f_k$ and the success probability $\eta_k$. For the subroutine $\mathcal{P}$:
\begin{equation}
\label{appen.eq.P}
\begin{aligned}
    & \eta_k = \sum_{ijk \in S } \text{tr} \! \left( \text{M}^{ijk}_\text{A}\, \text{M}^{ijk}_\text{B}\, \text{M}^{ijk}_\text{C}\, \rho_{k-1}^{\otimes 3} \right), \\
    & \varrho_k = \frac{1}{\eta_k} \! \sum_{ijk \in S } \!\! \text{N}^{ijk}_\text{A}\, \text{N}^{ijk}_\text{B}\, \text{N}^{ijk}_\text{C}\rho_{k-1}^{\otimes 3} \,\text{N}^{ijk\,\dagger}_\text{A}\, \text{N}^{ijk\,\dagger}_\text{B}\, \text{N}^{ijk\,\dagger}_C,
    \\
    & f_k = \langle \text{W} | \, \varrho_k \, | \text{W} \rangle,
\end{aligned}
\end{equation}
where $S = \{001, 010, 100\}$, $\text{N}^{ijk} = \ket{0} \! \bra{  {\rm W}^{ijk} } + \ket{1} \! \bra{ {\rm W}^{ijk} } X^{\otimes3}$ and $\varrho_k$ is a 3-qubit state.

For the subroutine $\bar{\mathcal{P}}$:
\begin{equation}
\label{appen.eq.Pbar}
\begin{aligned}
    \bar{\eta}_k & = \text{tr} \left( \bar{M}^{000}_A \, \bar{M}^{000}_B \, \bar{M}^{000}_C \, {\rm V}_A {\rm V}_B {\rm V}_C \, \rho^{\otimes 3}_{k-1} \right), \\
    \bar{\varrho}_k & = \frac{1}{\bar{\eta}_k} \bar{N}_A \,\bar{N}_B \, \bar{N}_C \, \rho_{k-1}^{\otimes 3} \,\bar{N}^{\dagger}_A \, \bar{N}^{\dagger}_B \, \bar{N}^{\dagger}_C, \\
    \bar{f}_k & = \langle W | \, \bar{\varrho}_k \, | W \rangle,
\end{aligned}
\end{equation}
where $\bar{M}^{000} = \Lambda^\dagger M^{000} \Lambda$, $\Lambda = H_1 H_2 H_3 \text{SWAP}_{1,3}$, ${\rm V} = X^{\otimes 3} + (\ket{000} - \ket{111})(\bra{000} - \bra{111})$ and $\bar{{\rm N}} = \left( \ket{+}\!\bigbra{W^{111}} - \ket{-}\!\bigbra{W^{000}} \right) \Lambda \, {\rm V}$, $\ket{\pm} \equiv (\ket{0} \pm \ket{1}) / \sqrt{2}$.

For subroutine $\mathcal{P}'$ we obtain
\begin{equation}
\label{appen.eq.Pprima}
\begin{aligned}
    \eta_k' & = \text{tr} \left( P^e_A \, P^e_B \, P^e_C \, \rho^{\otimes 2}_{k-1} \right), \\
    \varrho'_k & = \frac{1}{\eta_k} \widetilde{P}^e_A \, \widetilde{P}^e_B\, \widetilde{P}^e_C \, \rho_{k-1}^{\otimes 2} \, \widetilde{P}^{e \, \dagger}_A \, \widetilde{P}^{e \, \dagger}_B \, \widetilde{P}^{e \, \dagger}_C, \\
    f_k' & = \langle W | \, \varrho_k' \, | W \rangle,
\end{aligned}
\end{equation}
where $\widetilde{\text{P}}^{\text{e}} = \ketbra{0}{00} + \ketbra{1}{11}$.

Then for the stabilizer EPP we choose either subroutine $\mathcal{P}$ or subroutine $\bar{\mathcal{P}}$, and hence the state after the $k$ iteration and the state, the fidelity and the success probability are given by
\begin{equation*}
    \text{if } f_k > \bar{f}_k \Rightarrow \left\{
    \begin{matrix}
    \rho_{k+1} = \varrho_k \\ 
    F_{k+1} = f_k \\  
    p^{\text{succ}}_k = \eta_k
    \end{matrix} 
    \right. , \text{ otherwise } \Rightarrow  \left\{
    \begin{matrix}
    \rho_k = \bar{\varrho}_k \\ 
    F_k = \bar{f}'_k \\  
    p^{\text{succ}}_k = \bar{\eta}_k
    \end{matrix}
    \right.
\end{equation*}
and for the improved EPP we choose either subroutine $\mathcal{P}$ or subroutine $\mathcal{P}'$, and hence the state after the $k$ iteration and the state, the fidelity and the success probability are given by
\begin{equation*}
    \text{if } f_k > f'_k \Rightarrow \left\{
    \begin{matrix}
    \rho_{k+1} = \varrho_k \\
    F_{k+1} = f_k \\
    p^{\text{succ}}_k = \eta_k
    \end{matrix} 
    \right. , \text{ otherwise } \Rightarrow  \left\{
    \begin{matrix}
    \rho_k = \varrho'_k \\ 
    F_k = f'_k \\  
    p^{\text{succ}}_k = \eta'_k
    \end{matrix}
    \right.
\end{equation*}

The number of required resources, i.e., the expected number of initial states $\rho_0$ one needs to obtain a single copy of $\rho_k$, is given by
\begin{equation*}
    R_k = m_k \, R_{k-1} \, p^{\text{succ}}_k \sum_{j=1}^\infty j \left( 1-p^{\text{succ}}_k \right)^{j-1} = \frac{m_k}{p^{\text{succ}}_k} \, R_{k-1}
\end{equation*}
where $R_0 = 1$, $m_k = 3$ for the stabilizer EPP, and
\begin{equation*}
    m_k =
\begin{cases}
    3 & \text{ if } f_k > f_k'  \\
    2 & \text{ if } f_k \leq f_k',
\end{cases}
\end{equation*}
for the improved EPP.

In Fig.~\ref{fig:purification} we take as the initial state a $W$ state affected by local white noise, i.e., $\rho_0 = \mathcal{D}_p^{\otimes 3}\left( \proj{W} \right)$.

In Fig. \ref{fig:purification:d}, we analyze the performance of the protocols under noisy operations, meaning we consider local depolarizing noise acting before each iteration. In this case, we use the same equations as in the noiseless case but substituting $\rho_{k-1}$ by $D^{\otimes 3}_p (\rho_{k-1})$ in Eqs. \eqref{appen.eq.P}, \eqref{appen.eq.Pbar} and \eqref{appen.eq.Pprima}. Given a certain value of $p$ we can find the minimum required and maximum reachable fidelities with each of the protocols by numerical analyses.

\bibliographystyle{apsrev4-2}
\bibliography{Biblio_w_repeater.bib}

\end{document}